# Fashion Industry in the Age of Generative Artificial Intelligence and Metaverse: A systematic Review

Rania Ahmed[1, *], Eman Ahmed[2, *], Ahmed Elbarbary[3,4 *], Ashraf Darwish[5,6 *] and Aboul Ella Hassanien[7,*]

[1]Faculty of Computers and Artificial Intelligence, Modern University for Technology & Information, Cairo, Egypt.
E-mail: rmohamed@cs.mti.edu.eg
[2, 7] Faculty of Computers and Artificial Intelligence, Cairo University, Cairo, Egypt.
[2]E-mail : e.ahmed@fci-cu.edu.eg , [7]E-mail: aboitcairo@cu.edu.eg
[3]Faculty of Arts and Design, University of Applied Sciences in Jordan.
E-mail: a_elbarbary@asu.edu.jo
[4]Faculty of Applied Arts, Benha University.
E-mail: ahmed.elbarbary@fapa.bu.edu.eg
[5]Faculty of Science, Helwan University, Cairo, Egypt.
E-mail: ashraf.darwish.eg@ieee.org
[6]Egyptian Chinese University, Cairo, Egypt.
E-mail: ashraf.darwish.eg@ieee.org
[*]Scientific Research School of Egypt (SRSEG) www.egyptscience-srge.com.

**Abstract:** The fashion industry is an extremely profitable market that generates trillions of dollars in revenue by producing and distributing apparel, footwear, and accessories. This systematic literature review (SLR) seeks to systematically review and analyze the research landscape about the Generative Artificial Intelligence (GAI) and metaverse in the fashion industry. Thus, investigating the impact of integrating both technologies to enhance the fashion industry. This systematic review uses the Reporting Items for Systematic reviews and Meta-Analyses (PRISMA) methodology, including three essential phases: identification, evaluation, and reporting. In the identification phase, the target search problems are determined by selecting appropriate keywords and alternative synonyms. After that 578 documents from 2014 to the end of 2023 are retrieved. The evaluation phase applies three screening steps to assess papers and choose 118 eligible papers for full-text reading. Finally, the reporting phase thoroughly examines and synthesizes the 118 eligible papers to identify key themes associated with GAI and Metaverse in the fashion industry. Based on Strengths, Weaknesses, Opportunities, and Threats (SWOT) analyses performed for both GAI and metaverse for the fashion industry, it is concluded that the integration of GAI and the metaverse holds the capacity to profoundly revolutionize the fashion sector, presenting chances for improved manufacturing, design, sales, and client experiences. Accordingly, the research proposes a new framework to integrate GAI and metaverse to enhance the fashion industry. The framework presents different use cases to promote the fashion industry using the integration. Future research points for achieving a successful integration are demonstrated.

***Keywords***: Artificial intelligence, fashion design, generative artificial intelligence, metaverse, digitalization in Fashion, fashion luxury industry and fashion marketing and visualization of similarities viewer analysis.

# List of Abbreviations

| Abbreviations | Word |
|---|---|
| AI | Artificial Intelligence |
| AR | Augmented Reality |
| cGANs | Conditional Generative Adversarial Networks |
| CNN | Convolutional Neural Networks |
| DL | Deep Learning |
| GAI | Generative Artificial Intelligence |
| GANs | Generative Adversarial Networks |
| HMD | Head Mount Display |
| IoT | Internet of Things |
| NFTs | Non-Fungible Tokens |
| NLP | Natural Language Processing |
| NPCs | Non-Player Characters |
| SLR | Systematic Literature Review |
| VR | Virtual Reality |
| XR | Extended Reality |
| PRISMA | Preferred Reporting Items for Systematic Reviews and Meta-Analyses |
| P-GAN | Progressive Growing GAN |
| SRGANs | Super-resolution GANs |
| WGAN | Wasserstein GAN |
| MGCM | Multi-modal Generative Compatibility Modelling system |
| SF-GAN | Spatial Fusion GAN |
| PAN | pose alignment network |
| TRN | texture refinement network |
| FTN | fitting network |
| TSN | Texture Synthesis Network |
| AFN | Appearance Flow Network |
| PAINT | Photo-realistic Fashion Design Synthesis |
| AUC | Area Under ROC Curve |
| KID | Kernel Inception Distance |
| SSIM | Structural Similarity |
| IS | Inception Score |
| FID | Frechet Inception distance |
| NIQE | Naturalness Image Quality Evaluator |
| RMSE | Root-mean-square error |
| PSNR | Peak signal-to-noise ratio |
| MRR | Mean Reciprocal Rank |
| MSSSIM | Multi-Scale Structural Similarity Metric |
| F2BT | Fashion Fill-in-the- blank Best Times |
| LPIPS | Learned Perceptual Image Patch Similarity |
| SIS | Inception Score |
| D & G | Dolce & Gabbana |
| VOS | visualization of similarities |
| PRISMA | Systematic reviews and Meta-Analyses |

## 1. Introduction, Motivation and Contribution

Throughout history, the designs and fashions of clothes have been a manifestation of humanity's quest for aesthetic appeal. The dress has served as a tangible representation of human civilization. Fashion has long been closely linked to human identity, as seen using beads and jewelry in ancient cultures. Fashion in modern civilization has had a profound impact on several aspects of social life, influencing and mirroring shifts in political, economic, cultural, and social worlds [1].

Fashion can be succinctly described as the prevailing style or types of accessories and clothing that people wear by specific groups of individuals during a particular period. Significant disparities may be observed between the exorbitant designer clothes exhibited on the catwalks of New York or Paris and the commercially manufactured athletic apparel and urban fashion retailed in shopping centers and markets worldwide. Nevertheless, the fashion industry involves various aspects such as design, manufacture, marketing, distribution, retailing, promotion and advertising of different types of clothes, ranging from high-end haute couture and designer fashion to daily attire [2].

Fashion mirrors the transformation in cultural, aesthetic, political, economic, and social aspects of life. People employ fashion as a means of expressing their preferences and way of life to both them and the larger community. With the continuous increase in the global population, there is a corresponding rise in the need for clothing, fashion accessories, and other textile products[3]. Figure 1 illustrates the widespread use of the fashion industry from 2001 till 2024.

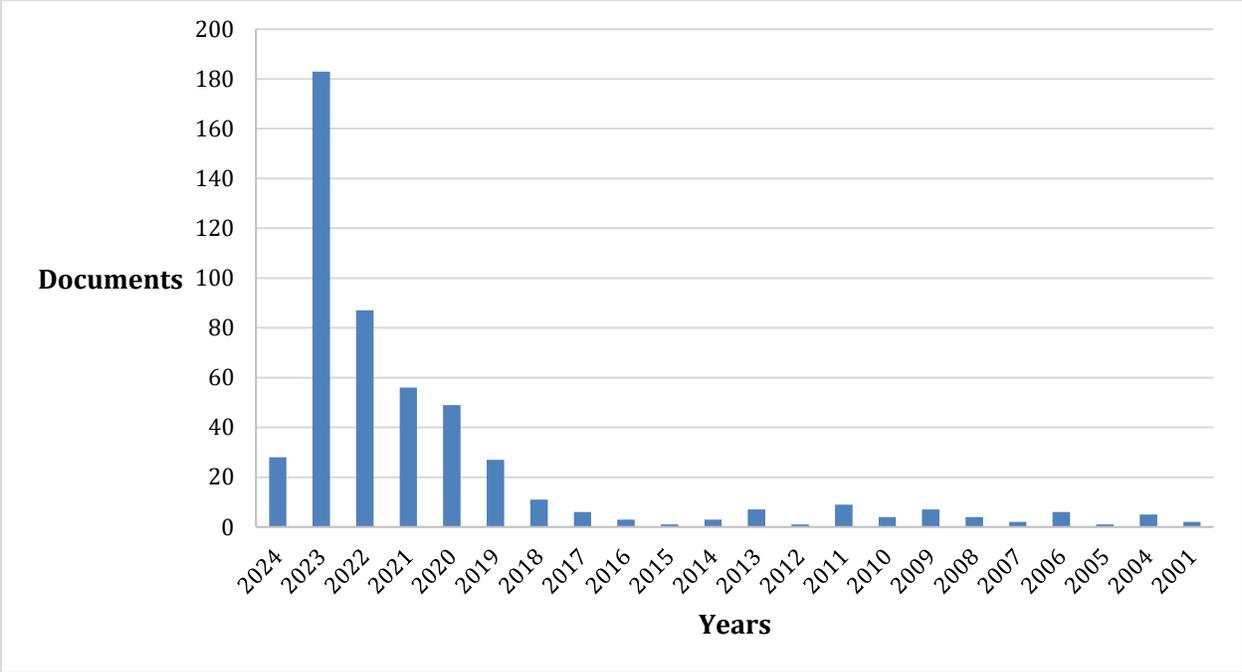

Fig.1.The popularity of fashion industry from 2001 to the beginning of 2024 based on the number of search interests.

Despite its origins in Europe and America, the fashion industry has now become a globalized and international industry. This entails the process of designing clothes inside a single nation, manufacturing them in elsewhere, and selling them in a third country. The fashion industry has historically stayed a major source for employment and continues to hold this position in the 21st century, contributing significantly to global economic production [4]. The fashion industry currently stands as one of the foremost economies worldwide, boasting an approximate value of up to 3,000 billion US dollars [5]. Figure 2 and Figure 3 show countries with the highest number of searches for the fashion industry.

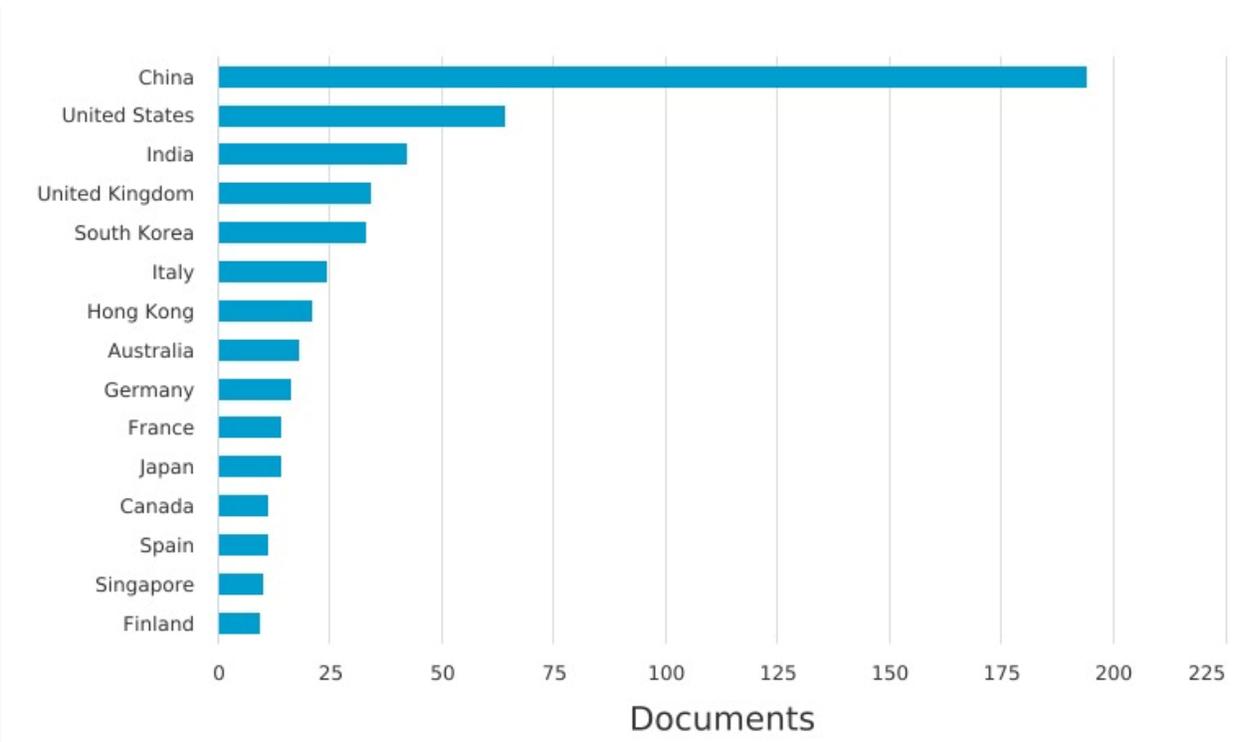

Fig. 2. Distribution of search interests in fashion industry.

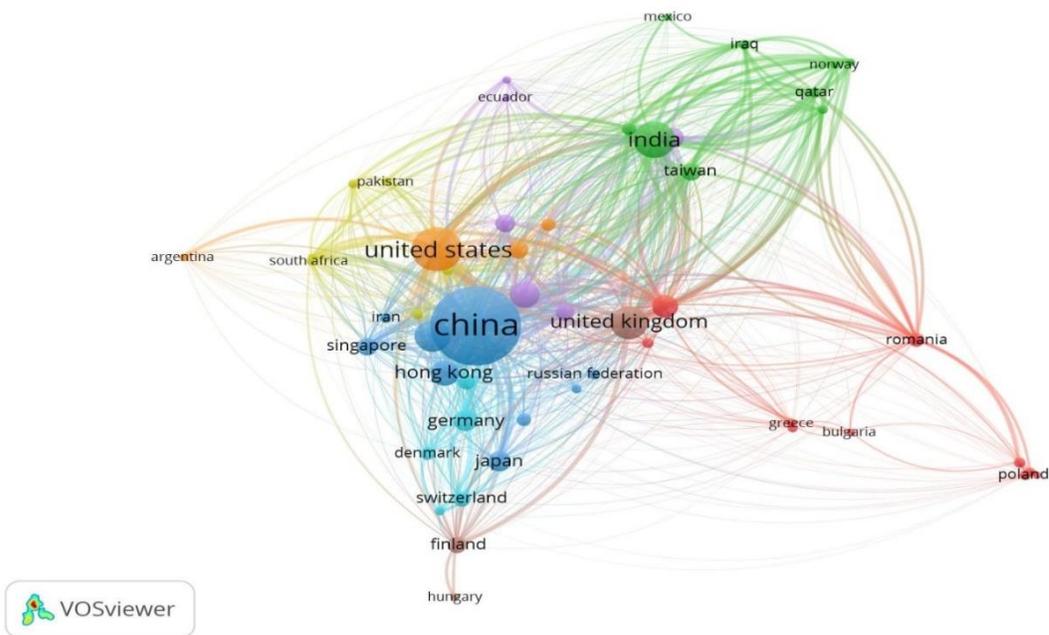

Fig. 3. Bibliographic coupling by countries of search interests in fashion industry.

Fortunately, the field of fashion studies has garnered growing interest in the computer vision, Artificial Intelligence (AI), and multimedia sectors in recent years [6] . GAI is a class of machine learning algorithms created to generate new, original content based on a set of input data. GAI is employed for a diverse range of tasks, encompassing the generation of text, graphics, music, and codes. AI-powered generative adversarial networks (GANs), a form of GAI, can carry out creative tasks that were previously believed to be restricted to people. These formidable machine learning algorithms can generate lifelike voice, images, and videos outputs [7].

The fashion industry greatly benefits from the implementation of GAI. By applying these strategies, online retailers can enhance the satisfaction of customers and expedite the introduction of innovative products to the market, while also reducing costs by:

- Expanding and individualizing fashion designs.
- Enhancing the inclusivity of body kinds by utilizing computer-generated models.
- Developing an automated system to enhance the digital user experience in online purchasing.

In the fashion industry, aesthetics and consumer satisfaction play significant roles in fashion design, while speed and creativity are essential. GANs provides a cost-effective method to create new product designs efficiently [8].

GAI is utilized to generate novel designs by using its training on various designer styles. The input can consist of sketches and the desired texture, and the model will provide the necessary designs.

Additionaly method that may greatly assist the fashion industry is known as the metaverse. The concept of metaverse has garnered significant attention in recent years with Mark Zuckerberg's announcement to rebrand his company from Facebook to Meta [9].

A recent study conducted by McKinsey reveals that the metaverse has the capacity to generate a staggering $5 trillion in value by the year 2030 [10]. This emerging virtual reality realm presents a significant growth prospect for various industries in the next decade, particularly the fashion industry. Numerous companies in this industry have already experimented with innovative branding strategies within the metaverse [11].
The proliferation of many platforms throughout time has caused ambiguity surrounding the term "metaverse," resulting in a lack of a universally agreed-upon definition [12]. However, a few recent contributions have proposed new conceptualizations of the metaverse based on distinct dimensions [13].

The metaverse can be conceptualized as technologically facilitated online collaborative shared spaces that are built upon 3D surroundings. These environments seamlessly combine physical and virtual realities, offering consumers highly immersive and social experiences. This enables the interaction between consumer-generated digital personas and the exchange of distinct digital assets [13] [14].
Mainly, people operating in this digital economy can engage, communicate, and trade goods and services digitally by means of avatars [15]. In this highly participatory society, exchanges take place using digital currencies and non-fungible tokens (NFTs), which serve as credentials of digital ownership [16].

The metaverse is constructed using fundamental technology components that form its underlying structure, including artificial intelligence (AI), augmented reality (AR), virtual reality (VR), mixed reality, extended reality (XR), cloud computing, and brain-computer interface. Blockchain technology, machine learning, computer graphics, and the Internet of Things [17] [18]. Interface devices are how consumers interact with and perceive a metaverse environment. The quality of the consumer's experience is determined by the degree of immersion, level of social interaction, and the extent to which the metaverse replicates the physical world [14]. The metaverse has a substantial influence on various economic sectors [19], especially those that require expanding chances for brand involvement[16], such as the fashion industry [20]. Aside from prominent tech corporations like Facebook and Microsoft, numerous fashion enterprises like Nike, Adidas, Balenciaga, and Gucci have commenced investment and delving into this realm via games, virtual storefronts, fashion shows, digital events, and so on [14]. Fashion brands are leading the way in the metaverse, with numerous luxury corporations already seizing the opportunities that arise from it and exploring novel avenues for consumer interaction [21] . The Metaverse Fashion Week held in March 2023 in Decentraland garnered participation from a diverse range of businesses and creative individuals, such as DKNY, Tommy Hilfiger and Dolce & Gabbanaand [22]. Gucci established the Gucci Garden within the Roblox virtual gaming universe, attracting a staggering 19 million visitors. Gucci also introduced the Gucci Virtual 25, which is considered the first virtual sneaker in the world. These sneakers, which are animated in 3D, were sold for $12.99 per pair [23].

In the last quarter of 2021, Ralph Lauren launched the inaugural digital apparel line, named "Ralph Lauren Winter Escape," exclusively on the Roblox platform [15]. The fashion industry's leaders have made significant strides in allocating resources towards digital investments in order to seize the financial, creative, and environmental advantages offered by the metaverse. By 2030, consulting organizations predict that the metaverse will account for 5% to 10% of the market value of fashion and luxury products, based on various projections. The market's value ranges from 360 to 380 billion euros [24].

The fashion industry has historically demonstrated adaptability and a willingness to adopt emerging technology. However, the driving forces behind its current shift towards the metaverse are distinct [25]. A possible reason is that fashion serves as a potent medium for individuals to express their individuality, personality, and inventiveness [26]. The Metaverse provides consumers with the chance to establish their digital persona, which can diverge from their physical identity. Furthermore, the metaverse has the potential to facilitate the advancement of fashion sustainability and extensive experimentation through novel aesthetics, socializing methods, and inclusivity for all fashion fans.
Research on the GAI and the metaverse's role in the fashion industry is currently limited, fragmented, and primarily centered around practical business applications.[27][28][23].

Recently, due to the swift advancement of deep learning, some researchers have started investigating the fields of art and fashion using the GAI and GAN framework [8] [29] [30]. Also, recent research has evolved with a specific focus on comprehending the metaverse and its application for businesses [14], particularly in the field of marketing [31] and consumer reception of this new technology [32]. Figure 4 shows the distribution of GAI with fashion industry citations per year over time, revealing a remarkable growth curve from 7 in 2013 to 177 in 2013, and it will grow even more in 2023. The exponential growth in citations demonstrates the rising popularity of this technology in the fashion industry and its significant role in addressing diverse optimization problems.

This SLR focuses on GAN as a crucial technology in the field GAI for the fashion industry. It seeks to transform fashion design and retail by using GAN, allowing for the development of cutting-edge and distinctive designs with incomparable efficiency and accuracy. This state-of-the-art technology enables the creation of realistic and varied fashion products, including clothing and accessories, by training models on large datasets of existing fashion images.

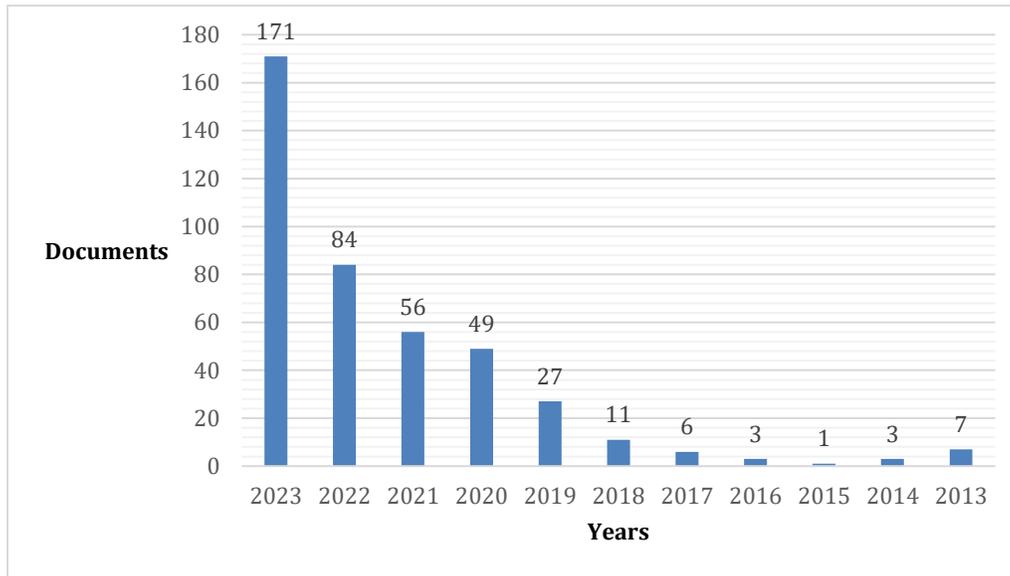

Fig.4. The annual citation counts of GAI with fashion industry.

Consequently, we aim to fill this research gap by carrying out a comprehensive examination of existing literature on the integration of the GAI and the metaverse in the fashion industry which is at the forefront of utilizing immersive environments. This is in addition to investigating the impact of making this integration. This SLR follows the updated PRISMA standards [33] for conducting and reporting systematic reviews and meta-analyses. This SLR study 188 relevant papers to review and explain critical themes related to applying GAI and Metaverse in the fashion industry. In addition, the SLR emphasizes their role in the fashion industry pointing out the potential challenges that may arise.

The main contributions of this SLR are summarized as follows:
- Providing a survey on using GAI in the Fashion industry to pave the way for the researchers and developers who are interested in improving and applying such techniques.
- Discussing the metaverse's role in enhancing the Fashion industry.
- Performing a SWOT analysis for GAI in the Fashion industry.
- Providing a SWOT analysis for the fashion industry's usage of the metaverse.
- Introducing the idea of integrating GAI with metaverse in the Fashion industry giving insights of the benefits and requirements and finally discussing future directions.

The paper's structure is presented in Figure 5, and it is arranged in the following manner. Section 2 provides an overview of relevant research. Section 3 outlining the research methodology employed in this SLR paper. Section 4 provides an explanation of fashion design in the digital era. Section 5 demonstrates the role of GAI in fashion design. Section 6 illustrates the relationship between fashion and metaverse followed by case studies in Section 7. Section 8 introduces the integration between GAI and the metaverse for the fashion industry with SWOT analysis performed for each technology. Discussion of results is provided in Section 9 while the SLR is concluded in Section 10.

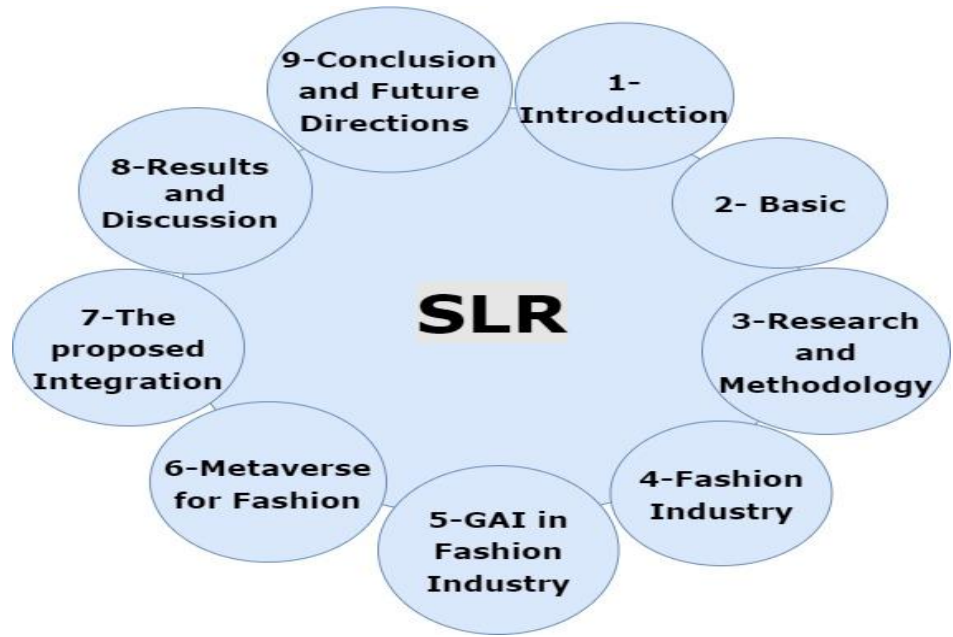

Fig. 5. The SLR structure

## 2. Basics and Background

The fashion industry, known for its constantly changing nature, is today being influenced by two powerful forces - the metaverse and GAI. In this emerging period of convergence between virtual and physical realities, the metaverse presents a vast platform for creative expression, social engagement, and immersive encounters. Simultaneously, GAI is changing the fashion industry by utilizing its capacity to analyze extensive datasets, create original designs, and improve personalized interactions.

This section presents fashion and AI-related works, including GANs theory and Metaverse which are relevant to the design of machine clothing.

### 2.1 Generative Adversarial Networks

GAI is a type of AI that learns from the training data to produce similar data. This is different from Discriminative AI that is trained on the training data with the target of classifying or recognizing test data. A type of GAI is Generative Adversarial Networks (GANs) [7], it includes two components: a generating model and a discriminative model. The generation model can be seen as a "forger" that attempts to deceive the discrimination model by creating counterfeit data. Conversely, the discrimination model can be seen as a discriminator that strives to determine whether the data come from real samples or simulated data generated by the generation model. Both models enhance their capability through ongoing adversarial learning. Accordingly, GANs learns in an adversarial way between the generator and the discriminator; the generator and discriminator compete in a way enabling the generator to improve its production to mimic the training data until it deceives the discriminator [34]. The generator usually starts by a random noise representing a low-dimensional latent space and train until it maps it to a high-dimensional latent space representing an image.

There are several variations of GANs. The standard GAN or Vanilla GAN proposed in 2014 consists of a generator and a discriminator, it uses Jensen-Shannon divergence for training. Conditional GAN (cGAN) [35], [36] in which both the generator and the discriminator are conditioned on additional information such as class labels. They are learned on labeled dataset and can generate images conditioned on specific attributes in contrast to unconditional GANs that generate random images.

The Deep Convolutional GANs (DCGANs) [37] in which both the generator and the discriminator are Convolutional Neural Network (CNN). Wasserstein GAN (WGAN) [38] uses Wasserstein distance instead of Jensen-Shannon divergence. It provides more stable training and improved output quality. Progressive GANs [34] start with low-resolution images and gradually increase the resolution during training. This technique allows the GAN to produce higher resolution images as well as to train more quickly than comparable non-progressive GANs.

Image-to-Image Translation GANs: they take an image as input and map it to a generated output image with different properties. For instance, an image-to-image GAN can be trained to take sketches of garments and turn them into photorealistic images of garments with different textures. Pix2pix [34] is an example is an image-to-image translation model based on paired data. DiscoGAN [39] and CycleGAN [40] are image-to-image translation models that do not need paired data. CycleGAN consists of two generators and two discriminators. One generator map from first domain to a second while the other learns the inverse mapping from the second to the First. Each discriminator learns the real and fake images of one of the domains. Cycle consistency loss is used for CycleGAN. To achieve synthesis realism in both geometry and appearance spaces, Spatial Fusion GAN (SF-GAN) employs two synthesizers: one for geometry and another for appearance.[41]. The former learns contextual geometries of background images and transforms and places foreground objects into the background while the latter adjusts the appearance factors such as the color, brightness and styles of the foreground objects and embeds them into background images. StarGAN [42] is a multi-domain image-to-image translation model that utilizes only one generator and one discriminator. StarGAN is a variant of cGAN that enables multi-domain image-to-image translation. It can generate images that exhibit different attributes such as color by conditioning on a specific target domain.

InfoGAN [43]extends the vanilla GAN by introducing an information-theoretic regularization term. It encourages the generator to learn disentangled representations by maximizing the mutual information between certain latent variables and the generated samples.

StyleGAN [44] is known for its ability to generate highly realistic and diverse images. It introduces a generator architecture that separates the model's latent space into style and content components, enabling fine-grained control over the generated images' attributes.

Text-to-Image Synthesis takes text as input and produce images that are plausible and described by the text like in StackGAN [45]. Super-resolution GANs (SRGANs) increase the resolution of images, adding detail where necessary to fill in blurry areas [44].

More variations exist with different purposes. Autoencoder-based GANs [46], [47], [48] combine the strengths of both autoencoders and GANs to achieve better performance in image generation tasks. In autoencoder-based GANs, the generator network is typically structured as an autoencoder, consisting of an encoder and a decoder, while the discriminator network remains unchanged from the standard GAN architecture.

Inverse GANs work inversely to GANs. Its objective is to map data samples from a high-dimensional space to a low-dimensional latent space. It consists of an encoder network and a generator network. The role of encoders is to map data samples from the input space to a lower-dimensional latent space, while the generator network reconstructs the original data samples from the latent space. Inverse GANs have applications in tasks like image-to-image translation, image super-resolution, and image editing, where the goal is to learn meaningful representations of data samples in a lower-dimensional latent space [49].

When multiple dimensions or components of the latent space in GANs correspond to distinct and interpretable factors of variation in the generated data, this is referred to as disentanglement of latent space. Stated differently, disentangled latent representations allow for more exact control over particular parts of the generated output by capturing distinct and relevant properties or qualities of the data. Controlling certain features is beneficial to manipulate an image producing different images with different attributes. More information about disentanglement can be found in [50], [51], [52]

## 2.2 The metaverse

The term "metaverse" was envisioned originally in the science fiction novel, Snow Crash, as a 3D virtual world inhabited by lifelike avatars [53]. In recent times, there has been a significant increase in the focus on a term used to describe a network of three-dimensional virtual environments that combines aspects of both digital and physical worlds [54]. According to a widely accepted description, the Metaverse is distinguished by several key features, including the enduring presence of identity and things, a communal setting, avatars, synchronization, three-dimensional spatial representation, interoperability, and an immersive and socially interactive user experience [55]. The Metaverse is widely regarded as an evolved iteration of virtual reality, providing users with a comprehensive and immersive environment that seamlessly blends physical and virtual worlds. This is achieved through the utilization of advanced technologies including artificial intelligence (AI), augmented reality (AR), virtual reality (VR), blockchain, non-fungible tokens (NFTs), haptic feedback, the principles of Web 3.0 and game engines among others. These technologies collectively enhance the user experience by providing a more integrated, realistic, and multi-sensory engagement [56]. Although there is still no definitive agreement on the definition of the metaverse, enterprises have already initiated efforts to investigate its viability as a marketing platform, aiming to leverage its potential [57]. As an illustration, the social media platform Facebook has undergone a rebranding process and is now known as Meta, with the aim of materializing the concept of the metaverse [58]. Similarly, Microsoft has made public its intention to acquire the game production business Activision Blizzard, a move intended to expedite its expansion through the creation of metaverse platforms [59]. Based on a recent study estimate, it is projected that the worldwide metaverse industry, encompassing hardware, software, and service sectors, would experience substantial growth, increasing from $16.2 billion in 2022 to $321 billion by 2027 [60].The fashion industry is not exempt from this phenomenon. In the year 2021, Nike acquired RTFKT, a firm specializing in the production of virtual footwear and digital artifacts. This strategic move was undertaken by Nike with the aim of expediting its digital transformation within the metaverse. According to recent research by [61], prominent brands including Nike, Dolce & Gabbana, and Tiffany have amassed a total of $232 million in revenue via non-fungible tokens (NFTs). Moreover, notable fashion labels including Gucci, Ralph Lauren, and H&M have initiated the establishment of virtual retail outlets on Roblox, an online gaming platform, with the intention of marketing and vending exclusively digital apparel and accessories. The evolution of the metaverse can be described in Figure 6[62][55].

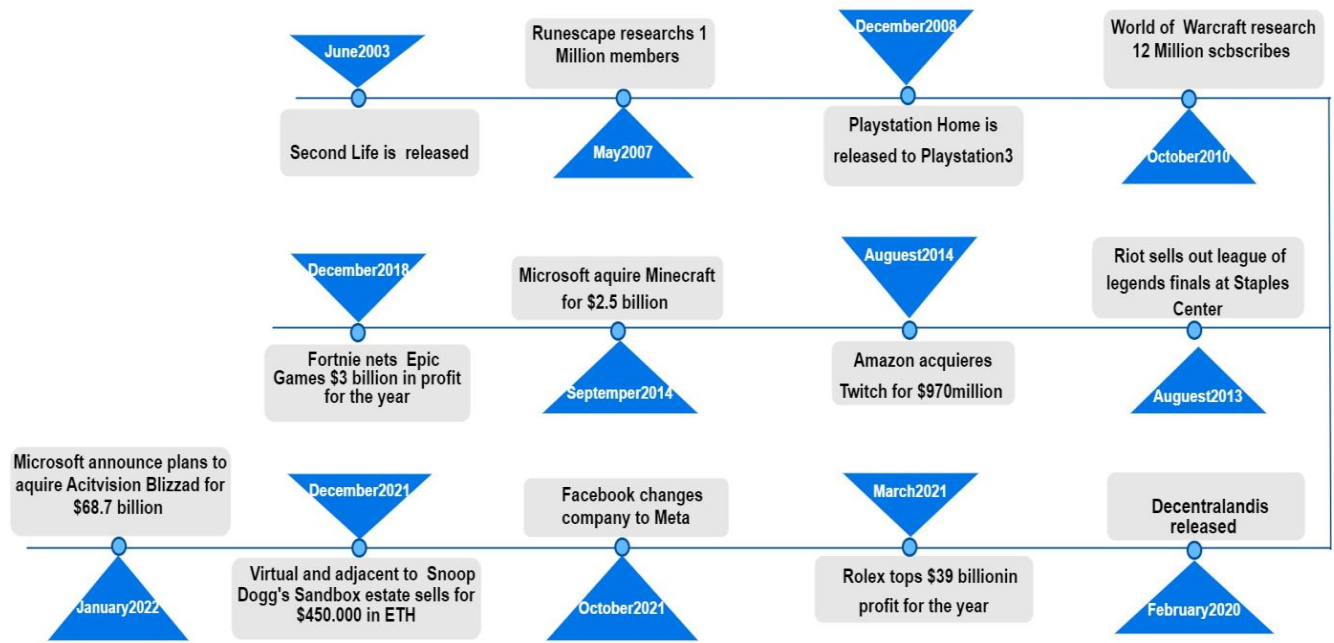

Fig. 6. The evolution of the metaverse.

The endeavor to replicate real-world experiences within virtual environments has been ongoing for several decades. Over the course of time, there has been a noticeable shift towards an augmented digital environment, which has had a profound impact on various aspects of our existence, including our consumption patterns in the realm of Fashion. The Metaverse has emerged as a lucrative opportunity for the fashion industry within the global pandemic, which has further intensified our reliance on virtual realities. In the contemporary era, virtual fashion has emerged as a prominent commodity, available for purchase through many channels. These channels encompass gaming platforms, digital photographs, augmented reality movies, and even novel revenue streams facilitated by non-fungible tokens (NFTs). This development presents a compelling prospect for the fashion industry to engage with hitherto untapped consumer groups and boost its financial gains. The Metaverse exhibits a notable array of advantages, effectively transforming a business that has traditionally relied primarily on in-person interactions into one that is increasingly intrigued and prepared to bridge the divide between the digital and physical worlds. In the realm of fashion design and branding, acquiring the ability to seamlessly integrate elements of reality and fantasy will prove crucial for professionals seeking to adapt to forthcoming trends and developments. In a similar vein, the field of digital design is expected to attract a substantial influx of creative production, necessitating that designers and companies possess a comprehensive understanding of effectively engaging their client base while simultaneously acquiring proficiency in digital tools. The fashion industry has been a prominent sector for numerous centuries, serving as a unique means and platform of communication. The implementation of technical and distributive methods to provide hyper-realistic 3D digital-only fashion has the potential to generate significant value and create several opportunities. This innovation has the potential to enhance digital interactions in everyday life and facilitate the development of more authentic digital environments. Consequently, it enables brands and designers to seamlessly transition into immersive and interactive settings, thereby creating new avenues to cater to untapped customer segments and markets [63].

The metaverse presents diverse prospects for enhancing the connection between the public and the fashion business. One notable opportunity in the fashion industry is the chance to participate in the Metaverse Fashion Week [64]. By employing this technology, the fashion industry can greatly enhance its creativity, interactions, and connection with the public, while also providing extensive customization options for digital items [28] [65]. The concept of personalization has always been a crucial aspect in the world of French "haute couture", where the focus is on crafting unique and original design products by hand. These principles can be applied to the metaverse through the utilization of immersive technologies [66].

Although there is a lack of not enough academic study on metaverse brand strategies, a few recent studies have examined certain branding techniques that employ fundamental metaverse technology to create positive brand value. As an example, Joy et al [23] studied the potential of cutting-edge technologies, including blockchain, NFTs, AI, and VR, to improve the perception of luxury businesses. Reay and Wanick [67] elucidated the way video games can delve into the convergence of gaming and fashion branding. Furthermore, Wanick and Stallwood [15] studied marketing tactics, including brand narrative, gamification, and social media marketing, in the metaverse. They conducted a case study on The Ralph Lauren Winter Escape to analyze these methods.

Meanwhile, certain researchers have emphasized the significance of essential technology in implementing metaverse brand strategies. Periyasami and Periyasamy [18] contended that avatars empower consumers to customize their digital depiction in the metaverse and enable marketers to engage with them in a manner akin to the physical world. In their study, Belk et al. [68] investigated the ownership challenges arising in the metaverse and proposed novel concepts of ownership, including fractional ownership and fractionalized property rights. In addition, many researchers have made efforts to develop theoretical frameworks for understanding the idea of metaverse by utilizing theories from marketing, advertising, and consumer behavior [69][70][71].

Several scholars argue that the metaverse offers unique chances to enhance consumer-brand connections, ultimately resulting in the establishment of robust brand equity[72] [73]. Hence, it would be beneficial to analyze fashion brands' metaverse strategies from the perspective of brand equity. Brand equity is a collection of valuable assets and potential drawbacks associated with a brand, including its name and symbol [74]. Brand equity is typically analyzed using two main approaches: a financial-based approach and a customer-based approach. Financial-based brand equity gives information on a brand's value for accounting purposes, while customer-based brand equity focuses on how consumers perceive and assess a brand and its products [75].

## 3. Research Methodology

In order to better understand how GAI and metaverse are used in the fashion industry, we performed a systematic literature review. This section discusses the methodology of the review including how the articles are selected and analyzed. The research method we adopted follows the guidelines pointed out in [76] and consists of the following consecutive phases:

- Identifying a specific number of relevant research questions.
- Extracting relevant keywords from the research questions to formulate appropriate queries.
- Specifying the databases in which the search has to be conducted.
- Applying preliminary filtering parameters, such as the search time range and the desired quality of the results, etc.
- skimming titles and abstract to exclude irrelevant articles and duplicates.
- Defining and applying detailed qualifying criteria throughout an in-depth review of surviving publications.
- Analyzing the remaining articles based on the research questions defined at the beginning of research.

This SLR paper is developed based on four main pillars GAI for the fashion industry, Metaverse for the Fashion industry, Integration of GAI and the metaverse for the Fashion industry, and future directions, as shown in Figure 7.

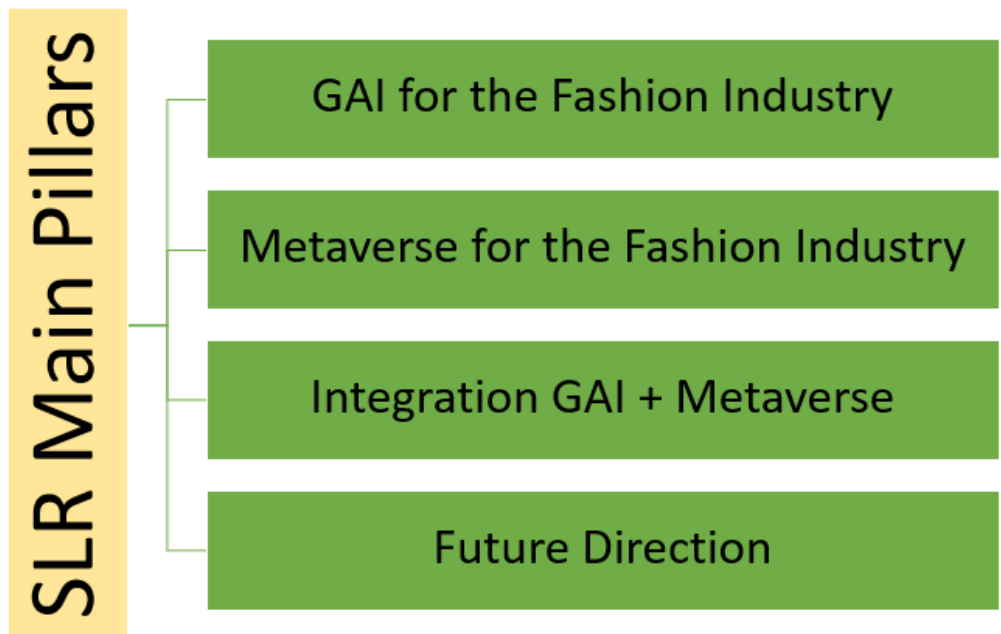

Fig. 7. The four main pillars of this SLR

The research methodology employed conforms to PRISMA [77] as shown in Figure 8.

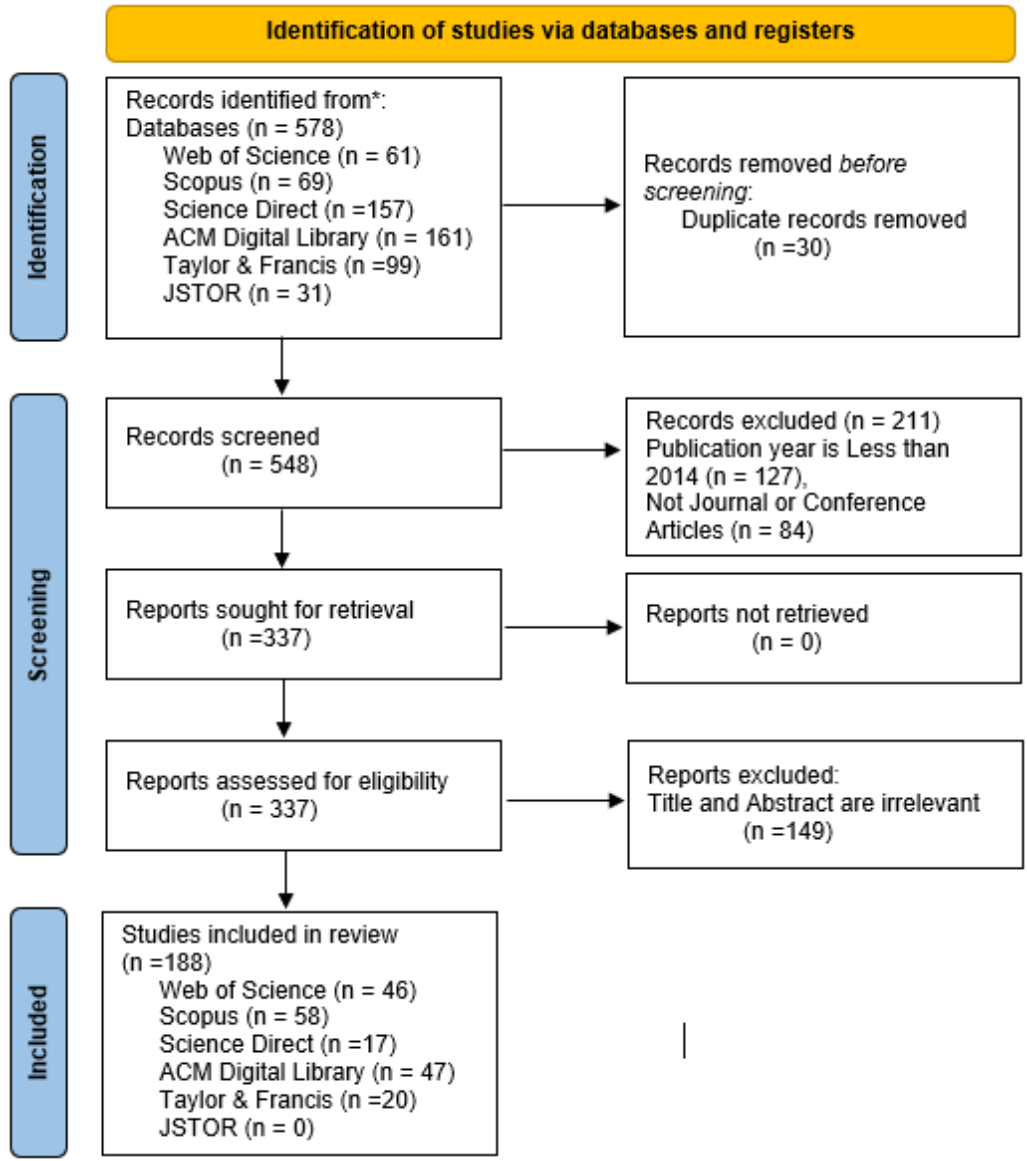

Fig. 8. Prisma flow diagram

**3.1 Identification and initially selection of research stage**

This stage has two steps first define the final search problems based on suitable queries, and next, extract the related documents from different databases, where this survey seeks to investigate the state-of-the-art GAI and Metaverse in the fashion industry. The research questions are the following:

- RQ1. What are the challenges facing the Fashion industry?
- RQ2. How can GAI be used to improve the fashion industry?
- RQ3. How metaverse can be used to improve the fashion industry?
- RQ4. What are the points of strength, weaknesses, opportunities, and threats of using GAI in the Fashion industry?

- RQ5. What are the points of strength, weaknesses, opportunities, and threats of using the metaverse in the Fashion industry?
- RQ6. How can GAI and the metaverse be integrated for fashion industry evolution?

The first research question aims to investigate how the fashion industry has been profoundly transformed by the digital age, which has revolutionized the way designers innovate, manufacture, and engage with customers. As regards RQ2 and RQ3 we referred to the GAI and metaverse presented in Section 2 and how the GAI and the metaverse holds the capacity to profoundly transform the fashion industry across all aspects, including design, production, retail, and customer experience, while RQ4 and RQ5 investigate the points of strength, weaknesses, opportunities, and threats of using the GAI and metaverse in the fashion industry. Finally, RQ6 studies the impact of integration of GAI and metaverse in the fashion industry. The research questions specified earlier have been transformed into appropriate queries, which were thereafter employed to inquire about the selected databases described in the next subsection. This SLR adopted the PRISMA process [78] for conducting systematic reviews and meta-analyses. This involves a three stages identification, evaluation, and reporting, in which we select a suitable group of papers for full-text reading.

This SLR used the following queries where the numbers correspond to those of the research questions:

> RQ1, RQ4, RQ6: ("generative" AND "artificial intelligence" OR "generative adversarial networks" OR "GAN") AND ("Fashion design" OR "Fashion industry" OR "Fashion image generation");
>
> RQ1, RQ3, RQ5: ("Metaverse") AND ("Fashion design" OR "Fashion industry" OR "Fashion image generation");

The paper focused on selecting publications from the following six main databases which are Web of Science, Scopus, Science Direct, Taylor & Francis, ACM Digital Library and JSTOR. Figure 9 and Figure 10 show a map based on bibliographic data when used integration between queries on the web of science and Scopus publications, respectively, by using visualization of similarities (VOS) viewer.

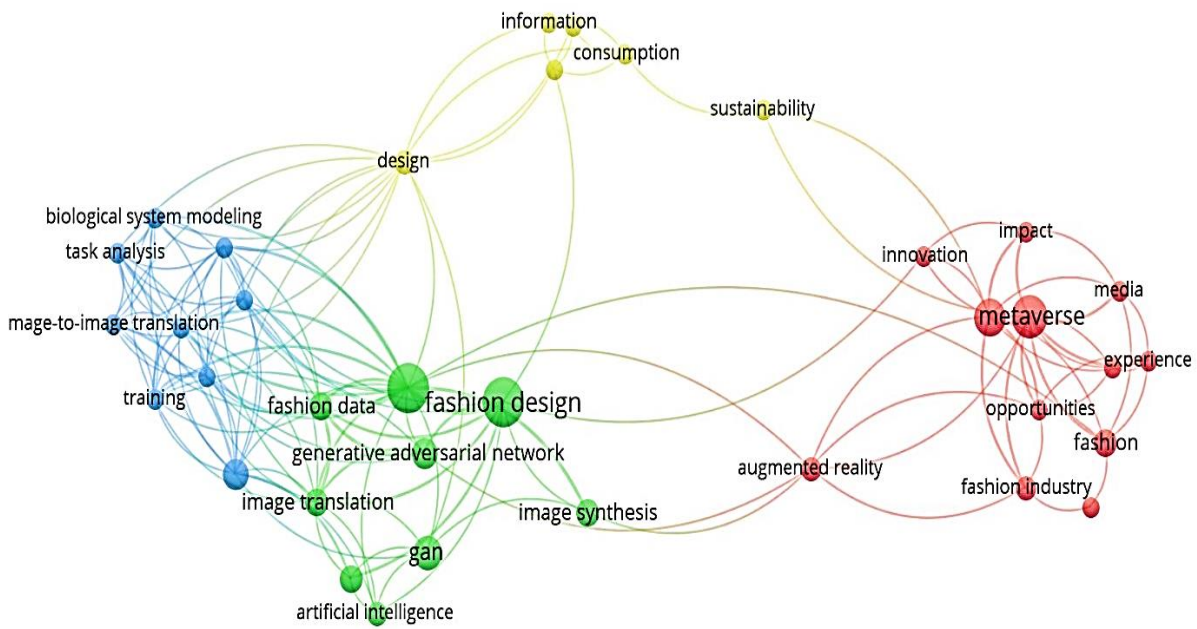

Fig. 9. The map of co-occurrence by all keywords in Web of Science

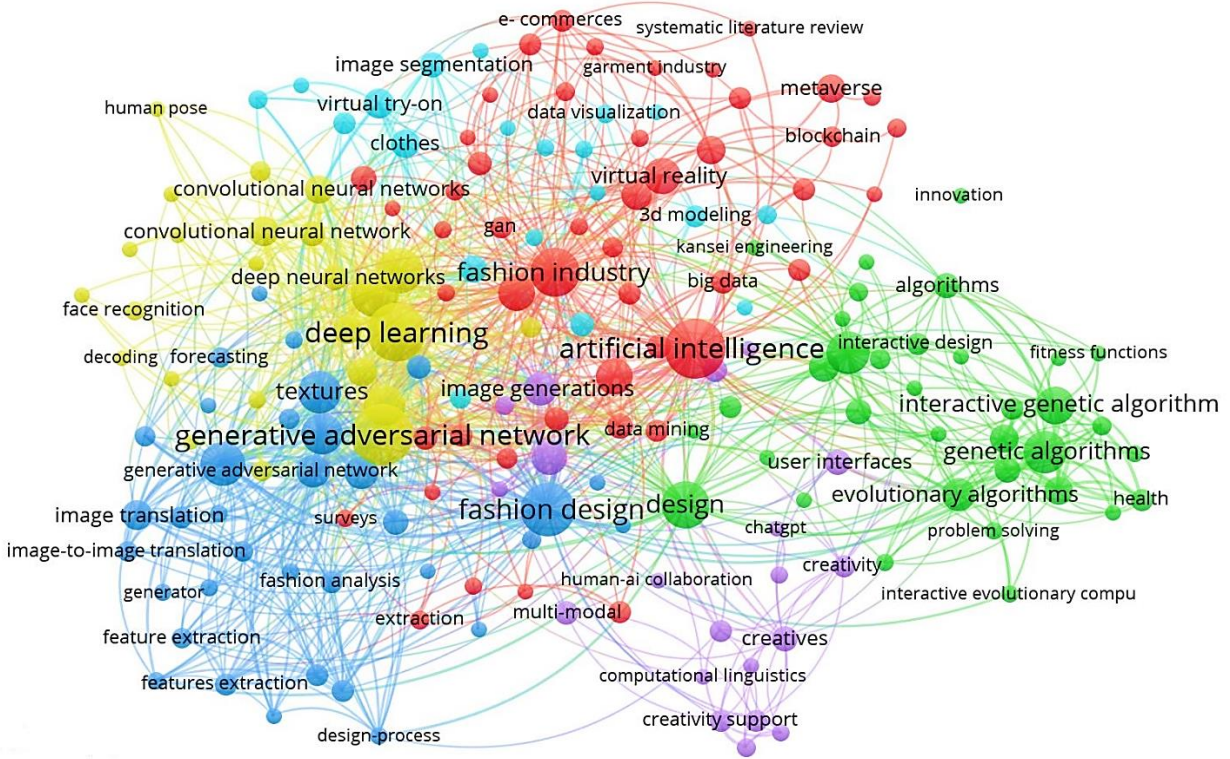

Fig. 10. The map of co-occurrence by all keywords in Scopus.

## 3.2 Evaluation stage

This stage consists of three screening steps. First, 211 duplicates and non-academic publications such as books, book chapters, abstracts, editorials and reports are removed from the 578 papers, leaving 337 reputable papers. Second, the title and abstract of each paper are scanned to select only those that have applied GAI and metaverse in fashion industry resulting in 188 related papers. The third step is to examine and study the full text of these 188 related papers.

Figure 11 shows the number of relevant publications per year. From 2014 to 2016, there is no relevant publications met according to the specified criteria mentioned above. It is shown that starting from 2017 until 2023, the number of publications is increasing. We have not included the number of publications in 2024 because at the time of publication the year is not finished.

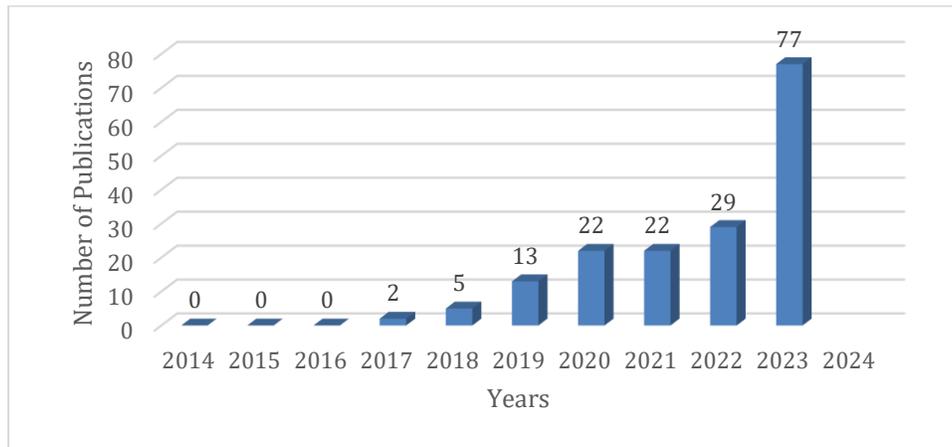

Fig. 11. The number of relevant publications per year.

## 3.3 Reporting stage

In this stage, the distribution of the selected papers by the six publishers is shown in Figures 12 where Scopus has the most published articles in this field distributed over the last ten years followed by ACM Digital Library then Web of Science with a comparable number. On the other hand, JSTOR had no relevant journal or conference articles. Then, the eligible papers are qualitatively analyzed and reviewed.

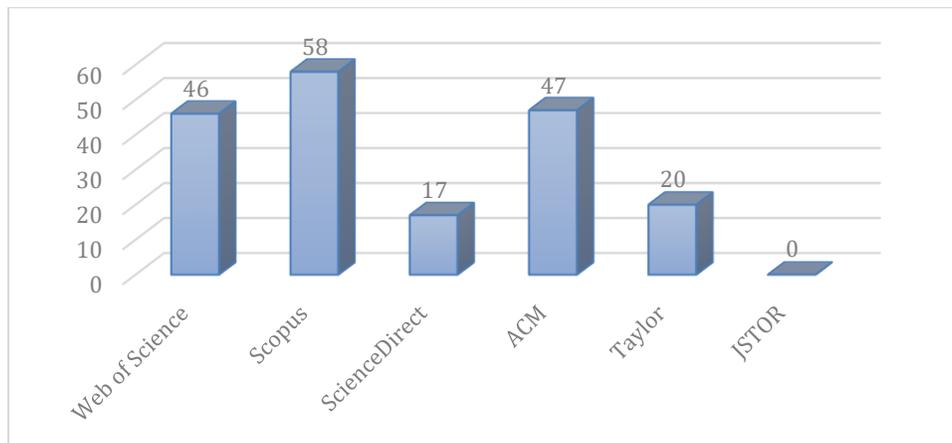

Fig. 12. Number of relevant publications of each of the used databases.

## 4. The Fashion Industry

The fashion industry is a versatile and multifaceted sector that includes the creation, manufacturing, promotion, and sale of clothing, accessories, footwear, and other associated lifestyle products [79]. It has a substantial impact on society, shaping not just how individuals portray themselves but also mirroring cultural, economic, and social trends. Clothing has played a crucial role in human civilization throughout history, providing practical, cultural, and symbolic functions. Over the course of history, fashion has transformed in reaction to cultural, economic, and technical advancements, mirroring changes in society values, aesthetics, and lifestyles. The fashion industry has a diverse range of participants, such as designers, producers, retailers, wholesalers, marketers, stylists, influencers, and consumers. Each entity plays a crucial role in the intricate ecosystem that fuels innovation, creativity, and commerce in this industry. The fashion industry is not homogeneous, but rather consists of several categories that cater to distinct consumers, demographics, and interests. The sectors encompass haute couture, luxury fashion, fast fashion, streetwear, athleisure, sustainable fashion, and others, each possessing unique attributes and catering to specific demographics [80]. The globalization of the fashion business has been enabled by advancements in transportation, communication, and trade. Designers derive inspiration from a wide range of cultures and work together with skilled craftsmen and manufacturers from around the world, while international brands extend their influence through extensive global supply chains and distribution networks [81]. Clothing not only plays a vital role in our everyday existence, but it also encompasses a substantial portion of the worldwide workforce, with over 300 million people employed in the sector as of 2017, and this number continues to rise consistently [82]. The fashion industry in the UK employs over 800,000 individuals across several sectors such as production, wholesale, retail, and services. This industry generates revenue exceeding 100 billion pounds [83]. The fashion industry in Europe employs almost 1.7 million individuals and generates over €165 billion in income [84]. The fashion sector generated approximately $1.5 trillion in global revenue in 2021[85]. Hence, it provides extensive backing to employment, particularly in economically weak countries, and substantially bolsters global economies.

Despite its appeal and inventiveness, the fashion industry faces numerous challenges, which include [86] Inadequate material and energy efficiency, deficient circular design, technological challenges, regulatory demands, internal stakeholders' pressures, budgetary constraints, insufficient human resources, ineffective management and leadership, absence of external partnerships, and issues connected to consumers. Table 1 presents a summary of the most recent articles with an emphasis on the challenges in implementing in economy.

Table 1. The challenges facing the fashion industry.

| | Challenges |
|---|---|
| [87] | 15 challenges encounter within the supply chain: 1. Inadequate control and legislation, 2. absence of support and governmental supervision, 3. mismanagement risk, 4. ineffective trash disposal, 5. inadequate backing from management, 6. short-term objectives, 7. inadequate funding available to the industry, 8. ineffective the framework was implemented, 9. inadequately trained labor force, 10. Resistance from employee to change |
| | 11. Inadequate preparation for the implementation and assimilation of Circular Economy and Industry 4.0,<br>12. Insufficient comprehension of Industry 4.0<br>13. use of materials as energy,<br>14. low-quality resources,<br>15. poor demand in the market |
| [88] | Categorize the challenges into two groups:<br><br>1. Internal challenges: Key factors that can impact an organization's operations include its product design, strategies and policies, lack of viable resources, internal stakeholders, technological difficulties, and financial challenges, collaborations.<br><br>2. External challenges: consumer-related challenges, economic challenges, social and cultural challenges, regulatory challenges, environmental challenges, and supply chain-related challenges. |
| [89] | Classify the challenges into seven distinct classes:<br>1. Environment,<br>2. Organizational,<br>3. Economy,<br>4. Social,<br>5. Institutional,<br>6. Supply chain,<br>7. Technological and informational elements. |
| [90] | Classify the challenges into two groups:<br><br>1. Lack of human resources and financial,<br><br>2. The intricacy of administrative responsibilities and laws. |
| [91] | Classify the challenges into three groups:<br><br>1. Challenges of operation,<br><br>2. Challenges of technical,<br><br>3. Issues related of customer. |
| [92] | Classify the challenges into two classes:<br><br>(1) Hard characteristics: financial challenges, regulatory challenges, circular business model- |

|   | related challenges, and pressures from stakeholders. |
|---|---|
|   | (2) Soft characteristics: challenges pertaining to consumers and green intellectual capital |

## 5. Generative Artificial Intelligence Tools in the Fashion Industry

In this section we provide an elaboration of the role of GAI tools in fashion design. AI-aided fashion design has an emerging interest because it eliminates tedious and time-consuming manual operations. Moreover, A fashion designer might draw inspiration and incorporate these produced images from the GAN structure. The application of GAI in Fashion can be categorized into four main categories, namely, virtual try-on, virtual garment display, style transfer and clothing matching. A virtual fitting room is a type of technology that allows customers to try on items virtually on their real-time images. Virtual garment generation and display is able to directly show the garment with design effects without the need to create a trial garment as done in the traditional apparel industry. More precisely, a user can define garment characteristics either through images such as sketches or text, then the garment is generated and displayed automatically. Style transfer refers to manipulating an item or (some items) from a current style to form a new one. Clothing matching refers to creating compatible sets of fashion items or choosing coherent fashion items that are in harmony with each other. Figure 13 depicts the four categories.

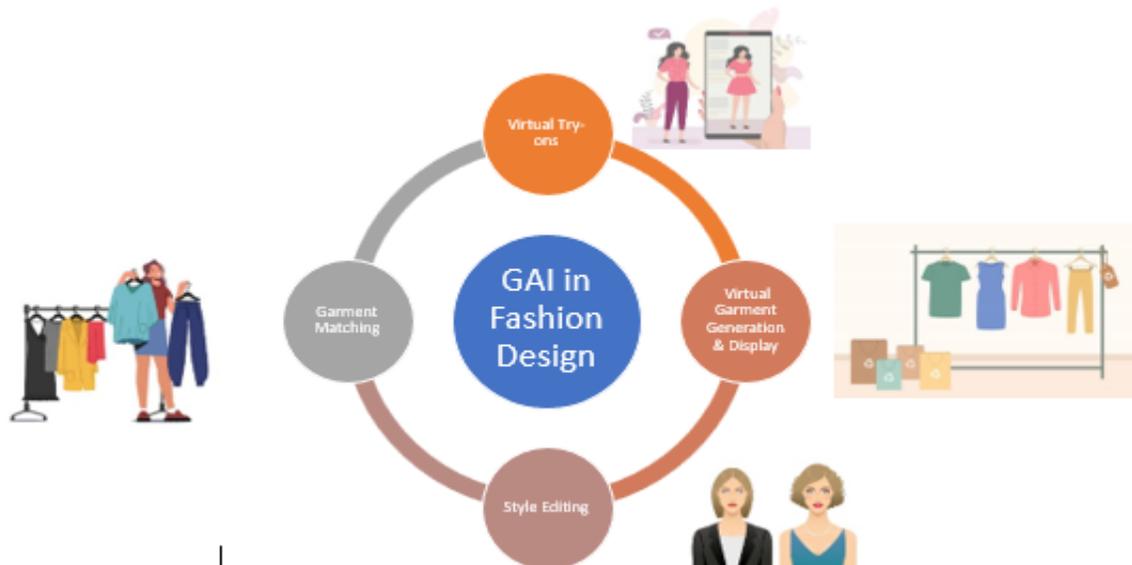

Fig.13. The categories of GAI usage in fashion industry

Regarding virtual try-on, one approach is human parsing and pose estimation. M2E-try [93] on Net takes as input a model wearing clothes and another person image. It then outputs the person image wearing the model's clothes. To achieve this, the suggested model uses three main components: pose alignment network (PAN) to align the poses between the model and the user; texture refinement network (TRN) to preserve the fine details and characteristics of the textures of the clothes and the fitting network (FTN) to fit the desired model clothes to the person. Similarly, PAINT [94] is a framework that produces a variety of multi-view fashion designs based on a human pose and texture examples of arbitrary sizes. It is divided into three stages. Firstly, the Layout Generative Network (LGN) is employed to convert an input human pose into a set of person semantic layouts. To synthesis textures on all altered semantic layouts, a Texture Synthesis Network (TSN) is used for the second stage. It performs an attentive texture transfer system to allow texture patches to be accurately expanded to the asymmetrical, uneven garment regions of the target fashion designs. During the third stage, by learning 2D multi-scale appearance flow fields, an Appearance Flow Network (AFN) is used for producing the images of fashion design that of several perspectives from a single-view remark.

Regarding virtual garment generation, FashionGAN [30] is an end-to-end virtual garment display method based on Conditional Generative Adversarial Networks. It takes a fashion sketch and a specified fabric image to the virtual garment image whose shape and texture are consistent with the inputs. In contrast to FashionGAN, GD-StarGAN [95] only requires the input of labels and textures in order to produce garment images that match the shape of the labels. GD-StarGAN is based on StarGAN but substitutes the original residual network of StarGAN generators with U-net [96] and uses an updated loss function. In [97] Progressive Growing of GANs (P-GANs) is applied for generation of clothing images for pattern makers. StyleMe, proposed in [98], can produce garment sketches that align with the designer's style through the use of a generative model based on deep learning. Additionally, the method allows for intelligent coloring of garment sketches by style transfer based on predetermined styles from actual fashion photos. The sketch generation stage consists of five modules: an encoder module, an adaptive channel feature normalization module, a generator with channel attention module, a loss function module and a discriminator with channel attention module.

An unsupervised network for creating and manipulating fashion styles is called UFS-Net [99] provides flexible editing and sketch-based generation despite using neither labels nor sketch in the training. In particular, a coarse-to-fine embedding procedure is first applied to incorporate both the actual clothes and the user-defined sketches into StyleGAN's latent space. Next, a feature fusion strategy is used to produce clothes using the characteristics given by the sketch. In [100], a fashion design based on artificial intelligence is proposed. This framework first introduces a sketch-generation module based on latent space for designing diverse sketches. Second, a rendering-generation module is presented to learn mapping between textures and sketches. The rendering-generation model develops a multi-conditional feature interaction module to achieve efficacy in synthesizing semantic-aware textures on sketches.

TAM GAN [101] first uses a pretrained language model known as TAM-BERT, which is fine-tuned on Tamil text data to encode semantic information, capturing the links between words and their meanings in context. Then, image synthesis is performed using DCGANs that gains the ability to convert textual semantic information into pixel-level representations of images.

StyleTrendGAN is an innovative deep learning framework used to create fashion goods. StyleTrendGAN utilizes a Dense Extreme Inception Network (DexiNed) for extracting sketches and Pix2Pix for converting the input sketches into new handbag models. InspirNET aims at generating fashion images with fine-grained texture details. This is done by introducing a controllable fine-grained texture disentanglement mechanism in a Generative Adversarial Network (GAN) framework. FashionGAN is presented with the goal of performing an image synthesis pipeline that can generate garment images based on a fashion sketch and a specified fabric image.

Style transfer involves altering the current style into another. In [102], segmentation is applied using Graphonomy [103] into 20 classes of all the body parts (background, hat, hair, gloves, sunglasses, top clothes, dress, coat, socks, pants, torso, scarf, skirt, face, right arm, left arm, right leg, left leg, right shoe, left shoe). A style encoder is trained to filter out region-specific style codes from the input image using convolution with the corresponding segmentation mask. Semantic region-adaptive normalization (SEAN) [104] is modified by adding a Resnet then squeeze and excitation network SENet [105] is used to transform clothes into the desired styles.

For clothing matching, OutfitGAN [106] consists of two main models, a GAN, which is responsible for producing high quality and realistic fashion items, and a compatibility network to measure compatibility among the items. An Attribute-GAN model is proposed in [107] for clothing match. The basis of the Attribute-GAN is using a generator under the supervision of an adversarial trained collocation discriminator and attribute discriminator. Multi-modal Generative Compatibility Modelling system (MGCM) [108] is another model used to create complementary pairings between the top and bottom of an outfit (e.g., matching shorts with a t-shirt). It is based on a conditional GAN architecture; it resolves an image-to-image translation task such that each article of clothing has a description that includes both verbal and visual explanations of the conditions. High-level similarities for the fashion items in the subdomain divided dataset (the top and bottom parts of the outfit) were highlighted by the researchers through the usage of compatibility limitations between items and between items and templates. Table 2 summarizes the techniques mentioned in each category. BC-GAN is specifically designed for batch generation, enabling the simultaneous synthesis of multiple visually collocated clothing images. his feature improves user practicality and convenience by providing a variety of clothing options to suit various situations and tastes. It uses a StyleGAN for the generator, InverseGAN for making disentanglement of features of the latent space and WGAN for the discriminator. The discriminator can learn compatibility by using a representation to push negative pairs further away and draw positive pairs closer in a common space using contrastive learning.

Table 2. A summary of the references' details including the reference number and the used application, dataset, basic GAN type and evaluation metrics.

| Application | Paper | Dataset | GAN Type | Evaluation Metrics |
|---|---|---|---|---|
| Garment Matching | MGCM [109] | • FashionVC[110]<br>• ExpFashion [111] | cGAN | • Area Under ROC Curve (AUC)<br>• Mean Reciprocal Rank (MRR) |
| Garment Matching | OutfitGAN [112] | Polyvore [113] and iFashion [114] | cGAN | • Inception Score (IS)<br>• Multi-Scale Structural Similarity Metric (MSSSIM)<br>• Compatibility Score |
| Garment Matching | AttributeGAN [115] | Collected from Polyvore | cGAN | • IS<br>• Human Evaluation |
| Garment Matching | BC-GANs [116] | Collected DiverseOutfit from Polyvore | • StyleGAN<br>• InverseGAN<br>• WGAN | • Fashion Fill-in-the- blank Best Times ($F^2BT$)<br>• Frechet Inception distance (FID)<br>• Learned Perceptual Image Patch Similarity (LPIPS)<br>• Human Evaluation |
| Virtual Try-on | M2E-Try On[93] | • DeepFashion [117]<br>• Multi-View Clothing (MVC) Women Top and Pants dataset [118] | cGAN | Human Evaluation |
| Virtual Try-on | PAINT[119] | Fashion-Gen dataset [120] | | • Inception Score (IS)<br>• FID<br>• LPIPS<br>• Structural Similarity (SSIM) |

| Garment Generation | UFS-Net [121] | GarmentSet [122] and Collected Fashion-Top | StyleGAN | - FID<br>- Kernel Inception Distance (KID)<br>- Classification Accuracy for attribute classification |
|---|---|---|---|---|
| Garment Generation | StyleMe [123] | Collected | Encoder Modules | - FID<br>- LPIPS<br>- Human Evaluations |
| Garment Generation | TAM GAN [101] | Oxford-102 dataset | DCGANs | F1-Score measure |
| Garment Generation | StyleTrendGAN [124] | Collected MADAME dataset | CGAN (pix2pix) | - IS<br>- FID<br>- KID |
| Garment Generation | P-GAN [125] | Collected from one brand | Progressive GANs | Human Evaluation |
| Garment Generation | FashionGAN [126] | - Collected from DeepFashion and E-commerce website. | cGAN | - Human Evaluation<br>- IS |
| Style Editing | [127] | - Collected from DeepFashion and webcrawling | cGAN | - Root-mean-square error (RMSE)<br>- Peak signal-to-noise ratio (PSNR)<br>- SSIM<br>- FID<br>- Naturalness Image Quality Evaluator (NIQE). |

Table 3 provides a summary of the applied GAN techniques found in literature for different fashion applications while pointing out the advantages and the disadvantages of each technique.

Table 3. A summary of each reference's pros and cons.

| Application | Paper | Pros | Cons |
|---|---|---|---|
| Garment Matching | MGCM [109] | - Using multimodal representation of inputs gets better results.<br>- Can be used to generate personalized compatible pairs | Limited or biased data affects the performance |
| Garment Matching | OutfitGAN [112] | - Flexible as it doesn't require a certain number of partial items to generate the complementary one.<br>- OutfitGAN provides fashion-on-demand, where customers can access customized and on-trend fashion items quickly and easily. | - Limited or biased data affects the performance.<br>- It may not always align perfectly with customers' preferences. |
| Garment Matching | AttributeGAN [115] | - Scalable.<br>- Use of semantic attributes. | - Limited dataset is used.<br>- More performance measures should be used. |

| Category | Method | Strengths | Limitations |
|---|---|---|---|
| **Garment Matching** | BC-GANs [116] | Ability to recommend multiple complementary outfit items | Need to be tested on more varied dataset. |
| **Virtual Try-on** | M2E-Try On [93] | • Able to preserve the characteristics of clothes and can transfer the garment from the model person to the target person while preserving the identity and pose.<br>• Using self-supervised learning with unsupervised learning to mitigate the lack of paired data of a person and its image wearing the desired outfit | Fails when there is a head pattern on the clothes. In this case, the networks treat this head pattern as head of the human instead of clothes region. |
| **Virtual Try-on** | PAINT [119] | The generated fashion designs exhibit high visual quality and realism, closely resembling real-world garments. | • PAINT can only synthesize limited clothing categories and fashion styles covered in the training dataset.<br>• may lack originality and fail to capture novel or unconventional fashion trends because it encourages the generation of fashion designs close to the training data distribution, limiting the model's intrinsic creativity.<br>• PAINT may experience visual quality degradation when input textures are complex or contain widely spaced stripe patterns or bright floral patterns. |
| **Garment Generation** | UFS-Net [121] | • It supports flexible editing and sketch-based generation without using labels or sketches in the training.<br>• Scalable<br>• Produces high quality designs. | • High Computational Capabilities.<br>• Large Datasets are required. |
| **Garment Generation** | StyleMe [123] | Can help professional and non-professional designers to create sketches and make style alteration in less time. | • Requires large high-quality dataset from fashion designers<br>• can only generate sketches of complete clothing, but some designers may only need sketches of local details of clothing such as buttons and collars.<br>• In the sketch generation stage, when the clothing image texture is dense or a variety of colors are interlaced, the sketch texture generated by the system is a low-quality pixel block.<br>• When the clothing's local details are complex and variable, the sketch generated |

| | | | by the system only has a basic contour.<br>• In the style transfer stage, when the specified reference clothing pictures include black and white, and the colors of generated clothing pictures are grey and white. |
|---|---|---|---|
| **Garment Generation** | TAM GAN [101] | Generate design by textual description using Tamil language. | • F1-Score may not capture all aspects of image quality and semantic alignment with the input text descriptions.<br>• Not tested on Fashion datasets. |
| **Garment Generation** | StyleTrendGAN [124] | • specialized for bag designs.<br>• can be customized. | Need to be tested on different datasets |
| **Garment Generation** | • P-GAN [125]<br>• FashionGAN [126] | Performing statistical analysis with ANOVA. | • Limited dataset is used.<br>• Not enough performance measures are used |
| **Style Editing** | [127] | Simple. | • Low quality.<br>• More datasets should be tested. |

## 6. The Uses of the Metaverse in the Fashion Industry

The Metaverse has the potential to impact the fashion industry, as it alters our ways of perceiving and engaging with fashion and style in various meaningful aspects as evidenced by the subsequent points:

❖ **Unlimited creativity and better design:** Within the metaverse, fashion designers are afforded the freedom to conceive extraordinary designs including garments and accessories, unencumbered by constraints related to material restrictions or the financial burdens associated with production.

❖ **A new way to reach global customers at low cost:** The cost of creating a virtual store on Metaverse is much lower compared to the expenses associated with creating a real store. Therefore, the metaverse fashion platform facilitates global accessibility, enabling prominent businesses to effortlessly connect with customers residing in many geographical regions.

❖ **Virtual runways:** Organizing fashion shows within the metaverse is an easy one process, thereby enabling designers to effectively exhibit their creative works. The virtual runways possess the characteristic of being geographically unrestricted, thus enabling global audiences to participate.

❖ **Inclusivity:** The promotion of diversity is a core tenet of metaverse fashion, as it allows for the representation of a wide range of body types, genders, and styles through avatars.

- ❖ **Instant Try-On:** Virtual fitting rooms provide customers with the opportunity to try on a curated selection of clothing and accessories within a virtual environment, known as the metaverse. This technology enables users to effortlessly experiment with different looks and appearances. It has been found to improve the overall online shopping experience and mitigate the frequency of product returns.

- ❖ **Sustainability:** The fashion metaverse effectively addresses the imperative of sustainability through its capacity to minimize reliance on physical production and mitigate waste generation. The production of virtual apparel necessitates distinct resource requirements compared to conventional garments.

**6.1 Case Studies Integration of Metaverse Marketing in Fashion Luxury Industry**

Currently, numerous entities in the fashion luxury industry are already leveraging metaverse technology to market their products. This section will discuss two case studies of the impact of metaverse on the fashion luxury industry.

**6.1.1 Adidas into the Metaverse**

Adidas virtual presence has expanded since 2010, providing customers with ongoing immersive experiences in the metaverse. Adidas established itself as a frontrunner in innovation by developing a 3D environment known as the Adidas Originals Neighborhood. This virtual replica captures the brand's 2010 Celebrate Originality advertising campaign on a worldwide scale. The AR Game Pack shoe comprised 5 Adidas Original sneakers, with each shoe featuring a distinct AR code inscribed on the shoe tongue. By using the phone camera to scan two QR codes, the consumer was granted exclusive entry to the virtual Adidas Originals neighborhood. Within the metaverse, there were contests that awarded participants with augmented reality-generated "trophies" upon successful completion, thereby making them eligible to win substantial rewards. As a result of the COVID pandemic, Adidas, like many other brands, had to shut down its physical retail outlets. Adidas implemented an AR feature on their store windows across Europe to commemorate the launch of their new environmentally-friendly Stan Smith sneakers. This function was made available through QR codes. The code was integrated into the website, thereby enhancing its energy efficiency through the reduction of software download requirements. The tagline "dare to change" and the presence of a "Shop Now" button enticed potential customers during their scanning experience, hence streamlining the purchasing procedure for them. Adidas acquired parcels of land in Sandbox and designated it as "adiVerse". The sandbox is a digital environment where users can possess, generate, and generate income from their experiences. This acquisition has provided the brand with new opportunities to engage in virtual brand activations, offer exclusive virtual items, and provide immersive fan experiences. Coinbase safeguarded the brand's digital assets. In an interview, Scott Zalaznik, the chief digital officer of Adidas, expressed that blockchain technology has the capacity to open limitless possibilities for engaging with our members. The establishment we are constructing with Web3 will result in novel prospects for collaborations, interaction through digital products, and a trajectory towards a more comprehensive future. Adidas collaborated with Bored Ape Yacht Club, NFT collector Gmoney, and Punks Comic for their inaugural foray into the metaverse. They acquired Bored Ape Yacht Club NFT #8774, transformed it into a metaverse persona named Indigo Hertz, and adorned it with a personalized Adidas-branded tracksuit. The compilation titled "Into the Metaverse" comprised a total of 30,000 non-fungible tokens (NFTs), each valued at 0.2 ETH. All of the aforementioned tokens were produced shortly after their initial release, resulting in Adidas generating around $22 million in revenue from their sales, considering the prevailing price of ETH during that period. To enhance the exclusivity of the NFT launch, a total of 20,000 units were made available to holders of Adidas Originals tokens, G-Money tokens, Bored Ape Yacht Club NFTs, Mutant Ape Yacht Club NFTs, and Pixel Vault NFTs through early access. The acquisition facilitated the realization of both digital and physical advantages from Adidas. NFT owners have exclusive access to exclusive merchandise releases and have contributed to the development of Adidas's next NFT-community offers. The metaverse project had a division into two distinct phases. The initial stage mostly focused on community initiatives, such as Prada's collaborative resource art project with Zack Lieberman and Hot Seats showcasing Bad Bunny. In 2022, Phase 2 of this system was introduced, enabling NFT owners to convert their tokens into tangible goods [128]. Adidas has introduced an online "Ozworld" collection, which is the world's first platform for creating avatars using personality-based AI, to promote its Ozworld collection. This metaverse platform is designed to facilitate the self-expression of Gen Z by providing them with the ability to create customized digital avatars. To remain faithful to their intended message, the avatars prioritize unique personalities over physical appearances. Users respond to a sequence of inquiries, which encompasses their preferred Ozworld footwear design, in order to explore their individual identity in greater depth. Subsequently, the platform generates a unique digital avatar, drawing influence from the engaging visual motif of the collection. Through a partnership with Ready Player Me, these avatars have the ability to navigate the internet and allow users to uphold a uniform virtual identity throughout various virtual environments, including more than 1500 metaverse applications and games. "Virtual Gear" was initially revealed on the Adidas Discord channel. It was only accessible to Phase 2 Capsule NFT holders and marked the brand's inaugural collection of interoperable digital wearables in the form of NFTs. The collection of 16 unique jackets was specifically developed for utilization by virtual avatars and may be accessed through a PFP dressing tool. The collection featured three exclusive wearables designed by creators and paid tribute to the artistic style of Bored Ape Yacht Club (BAYC), G-money, and Punks Comics. Adidas has collaborated with the esteemed luxury brand, Moncler, to introduce a novel marketing initiative titled "The Art of Explorers" within the metaverse. The campaign involved the integration of AI-generated explorers with mixed media art, accompanied by a digital immersion and a limited-edition collection of non-fungible tokens (NFTs). Moncler sent invitations to a variety of artists to create these explorers, taking inspiration from Moncler's "Art of Genius" exhibition held in London in February. The show also serves as the setting for the entire digital experience on the Moncler website, evoking the vibrant urban environment depicted in the show. This immersive encounter incorporates a diverse combination of auditory, visual, and three-dimensional animation, effectively immersing users within a simulated urban environment. The

collection features exclusive goods and 3D art, which can be purchased by buyers through digital billboards. Adidas, at the forefront of the industry, has declared a partnership with Bugatti in the metaverse to produce a unique football boot called X Crazyfast. To uphold a sense of exclusivity, a limited quantity of 99 boots were made available, each adorned with two slogans that have significantly influenced the core principles of both Adidas and Bugatti. These phrases, namely "Impossible is Nothing" for Adidas and "Create the Incomparable" for Bugatti, symbolize the company's unwavering dedication to excellence [129]. The next figure shows the evolution of Adidas into Metaversa.

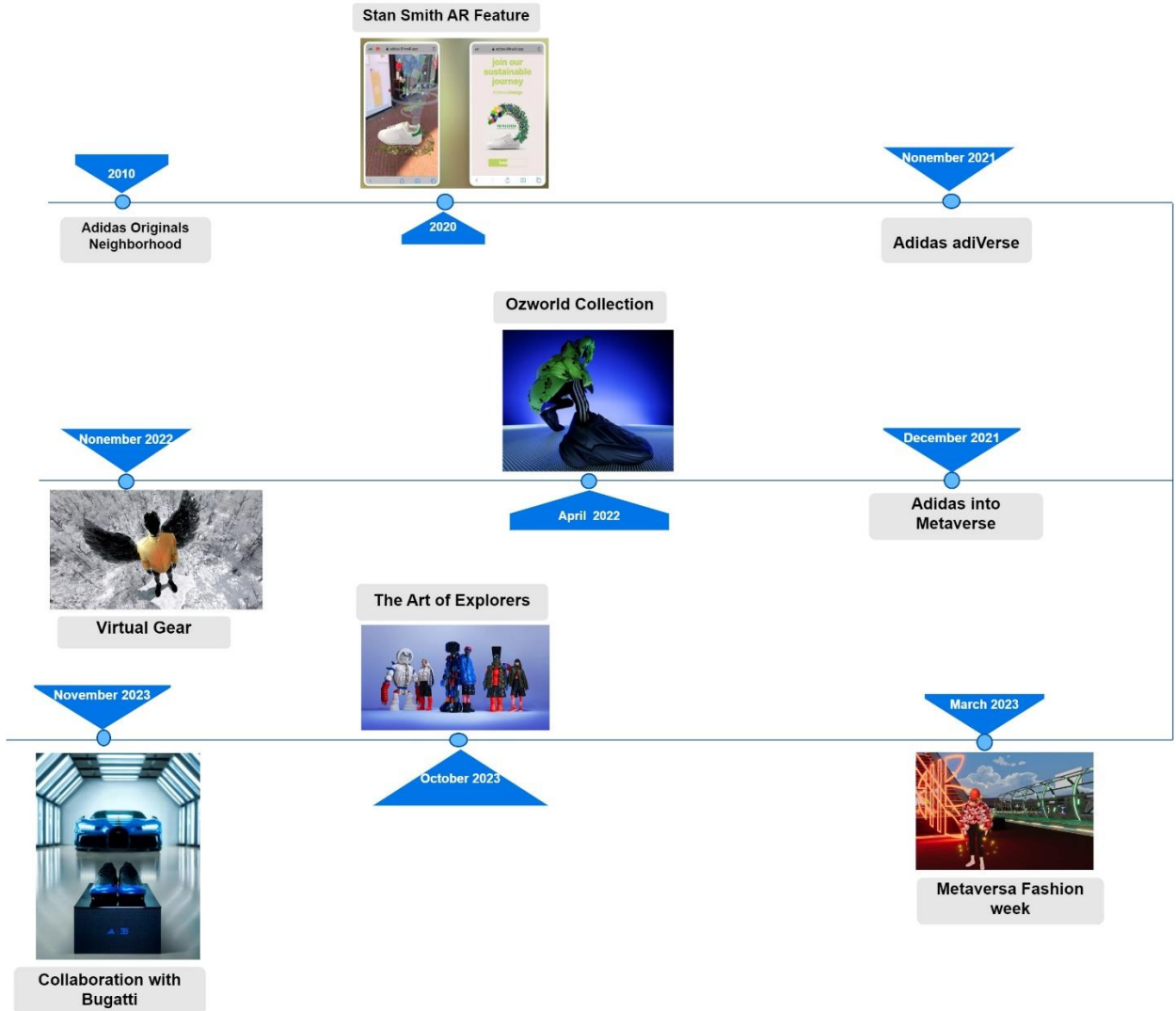

Fig.14. Adidas into Metaversa evolution.

### 6.1.2 Dolce & Gabbana

In October 2021, Dolce & Gabbana (D&G) also released a series of NFTs, although their approach differed from that of Adidas. Rather than targeting a broad consumer base, the Italian brand opted to develop a remarkably limited collection with only 9 NFTs. Five of the items have corresponding physical counterparts that were coupled with their respective NFTs. The brand achieved a remarkable accomplishment with this collection, setting a record for Fashion NFTs by generating $6 million in sales [130]. Several other digital-only creations that were made available for purchase achieved remarkable outcomes. The 'Impossible Tiara', adorned with seven blue sapphires and 142 diamonds, was purchased for $1.25 million, establishing it as the most expensive item in the collection [131]. Like Adidas, the digital assets provided assured entry to elite events, like the D&G haute couture shows or VIP tours of the workshops in Milan. Companies might choose to emulate either Adidas or Dolce & Gabbana's distinct approaches, while also prioritizing the establishment of a strong community and ensuring the authenticity of their product releases [132]. Figure 15 depicts an example of Dolce & Gabbana using the metaverse [133].

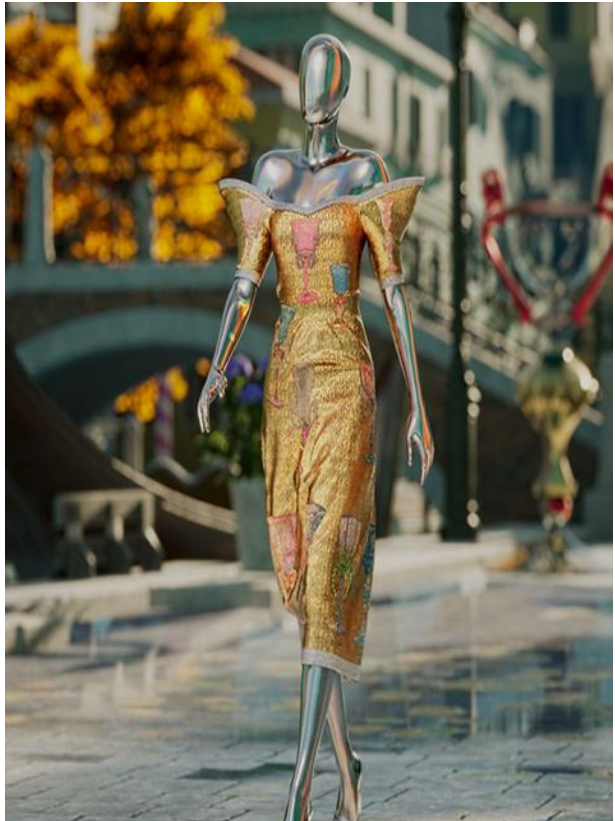

Fig.15. DOLCE & GABBANA X UNDX.

A team of analysts from Morgan Stanley suggests that luxury companies can capitalize on the growing metaverse market, which is projected to reach 50 billion euros ($57 billion) by 2030. This opportunity is expected to contribute to a revenue increase of over 10% for luxury companies, with a potential growth rate of 25%. The industry experienced a rise in EBITDA [134].

In the present day, the fashion and luxury business has embraced metaverse marketing, a new approach that has proven to be advantageous for firms.

The potential ideas for D&G future in the Metaverse can be summarized in the next points:

- **Virtual couture:** D&G has the capability to produce unique digital couture garments specifically designed for avatars.
- **Interactive experiences:** Envision participating in a virtual D&G fashion show or creating personalized D&G clothing within the metaverse.
- **Metaverse collectibles:** Limited edition non-fungible token (NFT) accessories or digital mementos have the potential to generate excitement among enthusiasts of the metaverse.
- **The next level of fashion shows:** The potential advancement of fashion shows lies in the possibility of livestreaming their physical shows within the metaverse, so enabling a broader audience to partake in the experience.

## 7 The Proposed Framework for Fashion Industry

As explained in the previous sections, both GAI and the metaverse are significant in the fashion industry. The existing section investigates the integration of both technologies to enhance the fashion industry. First, we perform SWOT analysis for each of the GAI and the Metaverse. Then, we investigate the opportunities of applying GAI in the metaverse. After that, use cases of GAI in the metaverse for enhancing the fashion industry is presented.

### 7.1 Generative Artificial Intelligence Analysis with Respect to Strengths, Weaknesses, Opportunities, and Threats

Figure 16 shows the SWOT matrix of GAI. GAI aims at learning the data patterns to be able to generate similar ones. This has many benefits in several applications like the fashion industry. Given examples of designs, GAI can generate similar designs as if they belong to the same designer. This saves time and effort for the designers in case of producing similar designs with different textures or fabrics. Moreover, the designer can try many variations of colors, texture, and fabrics and choose whatever he prefers for production. It is to be noted that before GAI, designers had to make the design then produce the model then decide whether it needs any modifications or not. Many fabrics have been wasted this way causing loss of time, effort, and money. With GAI, no need to produce models, virtual models' generation can replace the previous process. Only model production is performed when the designer makes his decision.

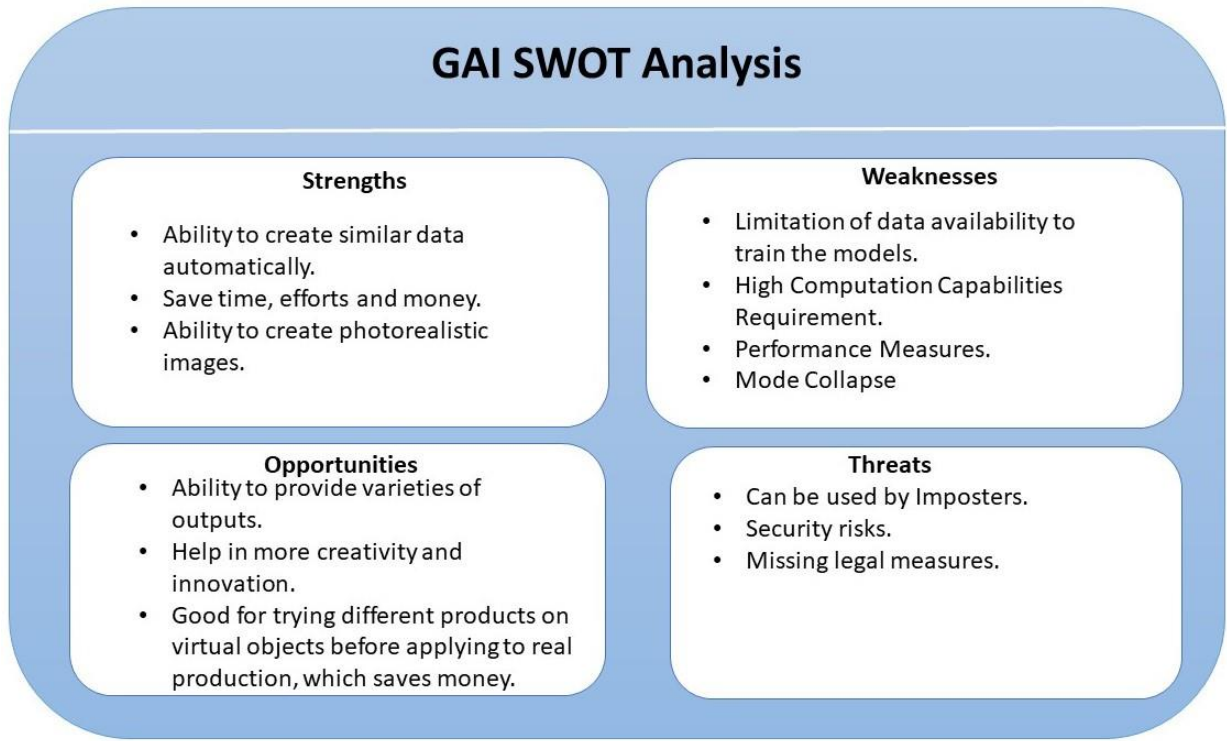

Fig. 16. GAI SWOT Analysis.

Still some problems remain for the application of GAI. GAI has to be fed by large amount of data to be able to learn properly to generate similar outputs. Without enough data, the outputs may not be similar to the original data. Sometimes, GAI can't reserve certain details of the texture and fabrics. Loss functions of the generator and the discriminator must be chosen carefully to generate outputs with the desired features. When a generator network fails to fully capture the variety of the training data distribution, it can produce a restricted set of outputs, a situation known as mode collapse in Generative Adversarial Networks (GANs). The generator tends to produce repetitive or nearly identical outputs, effectively collapsing to a single mode or a small subset of modes in the data distribution, rather than producing wide and variable samples. Another problem facing GAI is that it can be used by anyone. With the availability of famous designers online, imposters may be able to train a GAI model with a designer's products and generate similar designs. This represents a threat of using this technology. Thus, in order to preserve the innovations of the designers, the use of such technology needs to be protected by legal proceedings. Likewise, in order to prevent the theft of other designers' creations, legal measures must be taken to preserve access to online content.

**7.2 Metaverse Analysis with Respect to Strengths, Weaknesses, Opportunities, and Threats**

The Metaverse's numerous advantages are transforming a sector that has mainly depended on physical contacts, making it one of the most eager and prepared to bridge the divide between the digital and physical worlds [135]. The metaverse holds significant promise for fashion companies since it enables the development and customization of virtual retail environments, hence attracting new segments of consumers. Inside the metaverse, individuals have the ability to modify the virtual realm in order to highlight their brand's overarching goal, actively involve consumers inside the retail setting, and eliminate the distinctions between physical and digital domains. Companies may effectively engage customers by creating a compelling and immersive online experience, fostering a strong sense of reality and personal connection. This approach encourages users to repeatedly visit the virtual world. As an illustration, luxury brands are enhancing their presence on online and digital platforms by developing a metaverse to connect with younger customers. With stronger privacy regulations, technical advancements, and increasing expenses in paid digital marketing, companies are increasingly intrigued by the metaverse as a marketing tool.

In order to adapt to the future, fashion designers and businesses must acquire the essential skill of seamlessly integrating reality and fantasy. In the same vein, the advent of digital design is expected to attract a substantial amount of innovative work. Therefore, designers and businesses must possess the knowledge and skills to effectively connect with their customer base, while also becoming proficient in digital technologies.

Figure 17 depicts the SWOT matrix of the metaverse. The metaverse provides the users with an immersive experience in a virtual environment of different specs. The users are represented by avatars in the metaverse. In addition, non-play characters (NPCs) are created to provide more immersive experience to the users. Users can interact with each other through their corresponding avatars. It is to be noted that avatars can resemble the physical user in the real world to provide more immersive experience. An avatar has many characteristics, it has a face and body like the user, it can wear different clothes in different occasions, it can have different hair and clothes styles like in video games. Metaverse allows for imagination and innovations. Avatars can wear clothes from different eras and live in different eras through the creation of views and environments with suitable features. Similarly, users can play different roles in the metaverse. For instance, a user can be a doctor or a policeman or an actor. For each role, the user will wear different clothes that suit her/his role. Aside from entertainment, another benefit of the metaverse is that the user can live in different environments without leaving home. The user can also conduct training on different applications such as doing surgeries in a healthcare application. In this case, the user will train on virtual body as if he is doing the surgery in reality but without putting anyone at risk. A user can learn how to ride a vehicle in the metaverse instead of applying directly in reality. Although the metaverse has many benefits to users, still there are some problems facing its application. It needs real-time simulation to provide the user with an immersive experience. This, in turn, requires fast network connectivity. Moreover, virtual objects should resemble the real counterpart. Digital Twin technology is used for this purpose. In addition, the views of the environment should be photorealistic.

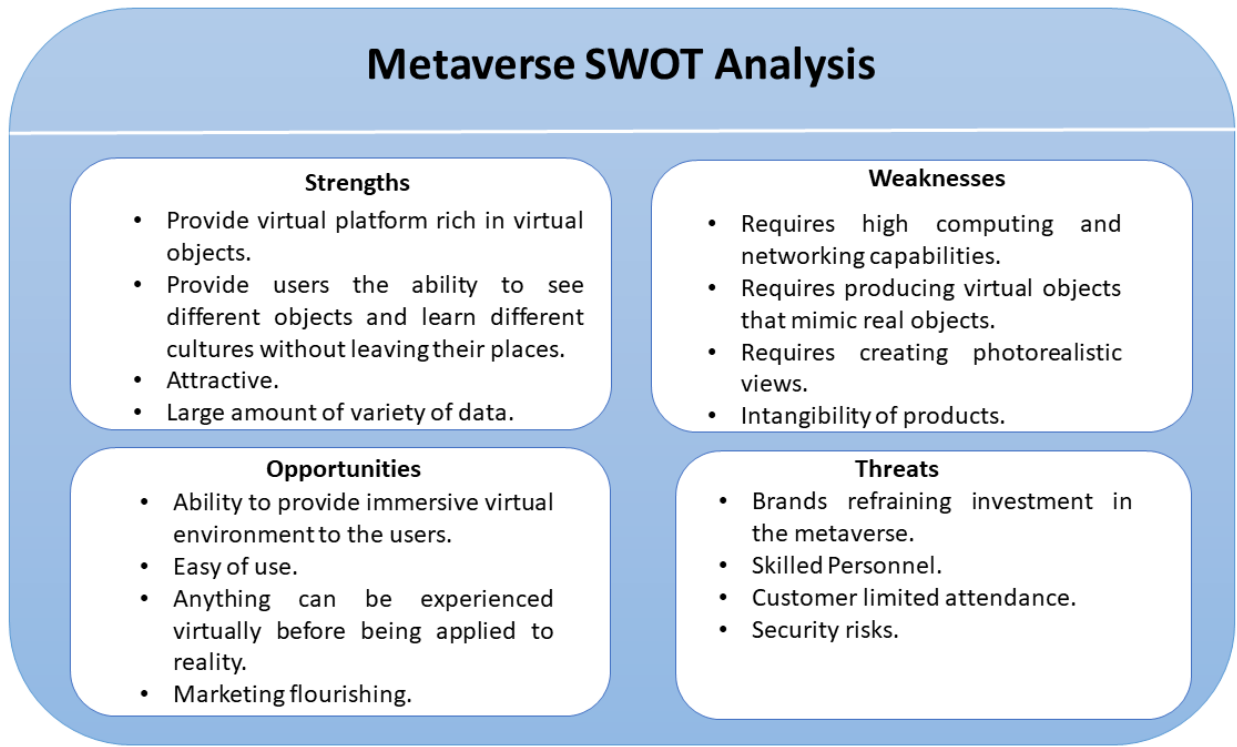

Fig. 17. The metaverse SWOT Analysis.

Among the threats that face the metaverse is the requirement of high computing and networking capabilities. Gaining access to the metaverse is a demanding task that necessitates substantial computational resources [72]. The infrastructure and processing capacities can provide a barrier to the fashion industry in the metaverse. This challenge specifically pertains to the ongoing or uninterrupted nature of anything. The metaverse is seeing technical advancements that require alignment between physical and virtual design elements [136]. The Metaverse is a technology that is currently developing, but its widespread use is hindered by intrinsic constraints in computer power [73]. Network connections need to be very fast to present an immersive experience to the user. Not being able to achieve these requirements badly affects the generated designs and the user experience, which, in turn, threatens the use of the metaverse.

The adoption of metaverse can be considered as a weakness point. There is increasing apprehension regarding the level of adoption and acceptance of the metaverse [137]. Despite the significant investment in infrastructure development, there are apprehensions over the manner and extent to which customers would visit and remain in the area. While there are assurances of increasing acceptance, it is reasonable for customers to seek information before joining. Several brands may choose to refrain from investing in this, and there are notable consequences for the creators. The client base is likely to diminish as brands decrease their investment in the metaverse due to limited attendance. Moreover, Obtaining the necessary technology infrastructure and skilled personnel to create fashion for the metaverse requires training that costs efforts and money. Another weakness point is "Intangibility of products", consumers should regulate their expectations while interacting with fashion brands in the metaverse. An important obstacle to consumer brand involvement in the metaverse is the intangibility of virtual fashion items. The intangibility of these commodities implies that consumers are unable to physically interact with or assess the quality of the products prior to purchase. The lack of tangibility might potentially result in unhappiness with the acquired things, particularly if they fail to meet the user's expectations [73]. An additional issue is the security risks that revolve with guaranteeing the security and privacy of personal data within virtual worlds. Personal data and digital assets held on a metaverse platform are susceptible to theft. As an illustration, the user's avatar data, including voice and video recordings, can be unlawfully accessed during platform usage, or an unauthorized individual can manipulate the avatar and exploit it. The metaworld possesses the capacity to amass a greater amount of delicate data compared to conventional systems, thereby posing a substantial risk to user privacy. Headsets equipped with active microphones can capture all conversations, while head-mounted displays (HMDs) with constantly operational cameras can record videos in private areas. Additionally, eye tracking technology can record the user's visual focus [138]. Security breaches in the metaverse encompass not just personal data, but also financial assets, such as cryptocurrency assets [139].

**7.3 Generative Artificial Intelligence and metaverse hybrid framework in fashion industry**

It can be seen that there is a mutual benefit between GAI and the metaverse with respect to the fashion industry. GAI can benefit from the large datasets available on the virtual platforms to train well and in turn help automate the creation of different clothes designs for the avatars in the metaverse. It also can produce photo-realistic views to provide the user with an immersive experience. Automation is necessary to cope with the dynamic nature of the metaverse. The user will be able to change her/his appearance and play different roles in the metaverse. This will be easy when using GAI. Additionally, fashion designers can benefit from the use of GAI in the metaverse to conduct virtual fashion shows that users can watch and interact with without leaving their places, hence, reaching global customers at a low cost. Users will be able to provide the designers with feedback which will enrich the fashion industry and help its improvement. This will leverage the experience of virtual try-ons in which users would gain the opportunity to try-on various designs to see what suits them and this will in turn helps marketing to revolutionize.

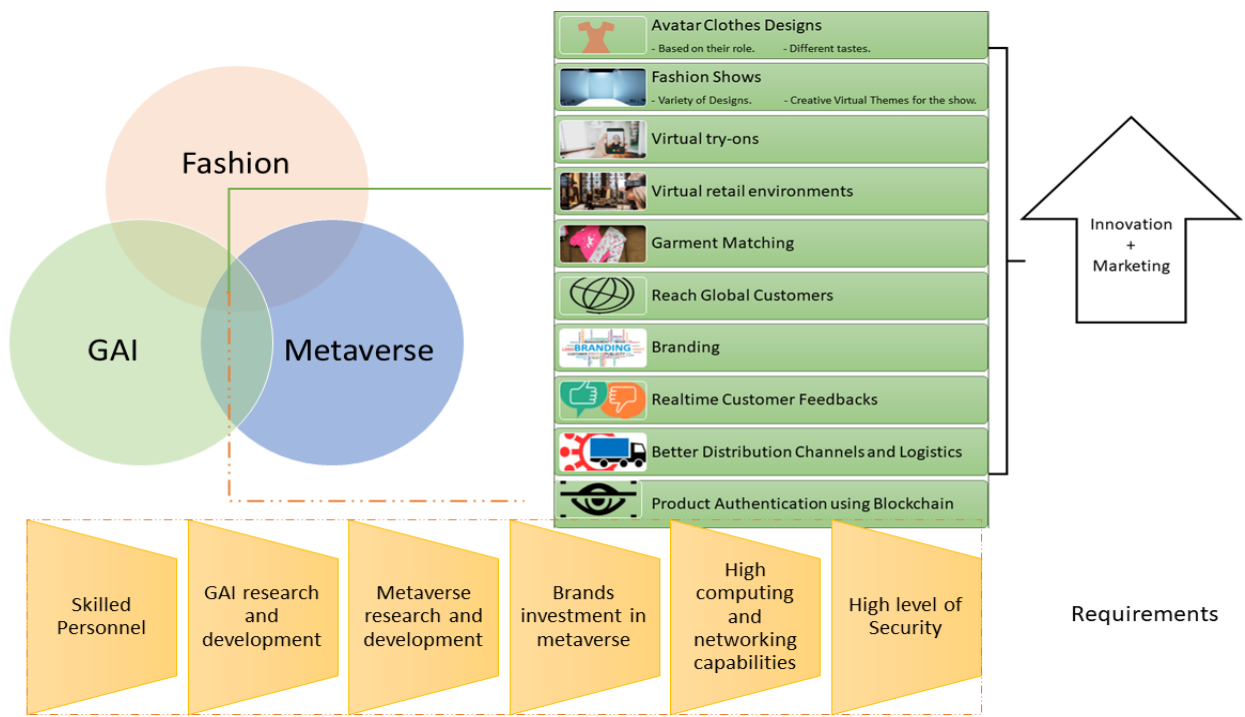

Fig. 18. The use cases resulting from the integration of GAI and the metaverse for the fashion industry and its requirements.

Figure 18 shows the use cases resulting from integrating GAI with the metaverse. The integration holds significant promise for fashion companies since it enables the development and customization of virtual retail environments, hence attracting new segments of consumers. Inside the metaverse, individuals can modify the virtual realm in order to highlight their brand's overarching goal, actively involve consumers inside the retail setting, and eliminate the distinctions between physical and digital domains. Companies may effectively engage customers by creating a compelling and immersive online experience, fostering a strong sense of reality and personal connection. This approach encourages users to repeatedly visit the virtual world. As an illustration, luxury brands are enhancing their presence on online and digital platforms by developing a metaverse to connect with younger customers. With stronger privacy regulations, technical advancements, and increasing expenses in paid digital marketing, companies are increasingly intrigued by the metaverse as a marketing tool.

In order to adapt to the future, fashion designers and businesses must acquire the essential skill of seamlessly integrating reality and fantasy. In the same vein, the advent of digital design is expected to attract a substantial amount of innovative work. Therefore, designers and businesses must possess the knowledge and skills to effectively connect with their customer base, while also becoming proficient in digital technologies. GAI researchers and developers play an important role in creating, maintaining, and updating techniques that help fashion designers to innovate and produce virtual designs on the metaverse. It will even help customers to match different clothes, try different styles and checks which suit them easily. This all will be accomplished at low cost, less time, less effort and less money spending than doing it in the real world.

The fashion industry has been a fundamental sector for millennia, offering a unique means and platform of communication. The ongoing progress of fashion brands in established virtual worlds (including virtual fashion, games, NFTs, digital skins, etc.) such as The Sandbox, Decentraland, Roblox, etc. indicates various significant aspects of company growth in the metaworld [140][31]. Table 4 explains the opportunities offered by the integration of GAI and the metaverse for the Fashion industry. Transferring a brand from physical to new virtual environment to enhance its value, engaging with customers through virtual products, and strengthening brand recognition by integrating gaming elements and connecting the virtual and physical worlds.

Table 4. Opportunities offered by the integration to the fashion industry.

| The field of development | Characteristics |
|---|---|
| Avatar virtual fashion and digital skins | Creating immediate and sustainable income by modifying the visual representation of avatars in games and online platforms. A significant proportion of consumers in various nations, such as the US and China, exhibit a strong inclination towards the creation and customization of products, creating digital personas and obtaining virtual possessions. GAI can assist in generating various virtual designs from which users can choose their favorites through the metaverse. |
| Virtual Fashion Shows | GAI enables the production of creative fashion shows in the metaverse. The themes and designs of the show help to attract customers and provide them with an entertaining experience. Moreover, customers from all over the world are able to |
| | present their feedback and choose to purchase among different products. |
| **Virtual Try-Ons** | Metaverse users will be able to try-on different designs virtually with the help of GAI without leaving their places. |
| **Virtual Retail Environment** | The ability to develop and customize virtual retail environments, hence attracting new segments of consumers. Inside the metaverse, individuals can modify the virtual realm in order to highlight their brand's overarching goal, actively involve consumers inside the retail setting, and eliminate the distinctions between physical and digital domains. GAI can help in creating virtual environments that mimic physical counterparts or even build innovative ones with better order and organization with various themes that attracts the users. |
| **Garment Matching** | The metaverse users will be able to perform matching between different garments virtually using GAI and choose the preferred ones easily. |
| **Reaching Global Customers** | Metaverse plays an important role in reaching customers all around the world. Hence, it provides a great opportunity for marketing different products. With innovative designs achieved with the help of GAI in short time, more customers will be attracted to this market which in turn will help in flourishing the market in the real world. |
| **Branding** | By integrating a brand into a virtual environment, it can enhance its visibility and appeal to customers. This can be achieved by offering virtual products, engaging with potential and existing customers, and utilizing gamification to strengthen the brand's reputation. Additionally, this approach can create a harmonious relationship between the virtual and real worlds. |
| **Customers feedbacks** | Consumer data availability, both in terms of its quality and quantity. Abundant prospects for conducting experiments to get data on consumer responses to novel product concepts or ideas. |
| **Distribution channels and logistics** | By utilizing Non-Fungible Tokens (NFT), we aim to dismantle the barriers that separate the physical and virtual worlds. This act as the bridge between the metaverse and the physical world enabling easier purchasing process to the customers after choosing their preferred products, moreover, it enables the customers to |
| | examine the products physically before purchasing the product giving them more confidence in the metaverse advertisements. |
| **Product Authentication** | NFTs provide the procurement, ownership, and exchange of distinct virtual artifacts that are authenticated by blockchain technology. NFTs enhance brands' product portfolios by introducing virtual offerings and fostering greater engagement within the virtual and real world.<br><br>NFTs as digital twins that securely record details on the historical background, genuineness, and possession of a physical or virtual item. This is particularly crucial within the luxury segment to combat the proliferation of counterfeit goods. NFTs function as "loyalty tokens" that provide consumers with added advantages, like exclusive access to new NFT releases and tangible merchandise. |

Some requirements are essential for the success of the proposed integration. Updated infrastructure that provides high computing capabilities and networking is necessary, without it, users will get bored and won't be attracted to use these technologies. There is a great need to invest in GAI research and development to enhance the models to create outputs with high quality and speed. Similarly, investments are need in the metaverse research and development to improve the technology and make it easy to use by different users. Creating avatars and overlaying different designs on them should be done easily. In addition to taking the body measurements and pose into consideration. Also, organizing virtual events needs to be easily done by different users. Fashion brands may have some concerns in using these technologies but their use to these systems is inevitable in the future. Awareness should be spread among different brand explaining to them how they can benefit from the use of the metaverse and GAI for making better products and enhancing their marketing capabilities. This is in addition to providing them with the availability to try these technologies and test their impacts. Moreover, without skilled personnels who are able to use both GAI and the metaverse, the integration would not succeed. Hence, proper training on the usage of both technologies is required. Security risks present a primary threat in both technologies. Accordingly, legislative and legal measures need to be prepared to protect the fashion designers' intellectual properties and the metaverse users' privacies. This is inevitable since security risks would cause users to refrain from using both technologies despite their benefits.

## 8   Results and Discussion

This section provides a summary of how the article addresses the analytical questions posed in Section 3.2. Table (5) shows the answers of the stated research questions of this paper.

**Table (5)** the answer to the research question mentioned an the beginning of this paper.

| Question# | Research Question | Answer |
|---|---|---|
| Q1 | **What are the challenges facing the Fashion industry?** | The article provides an explanation of the main technologies used in fashion design. This can be found in Section 4. |
| Q2 | **How can GAI be used to improve the fashion industry?** | The basics of GAI is presented in Section 2 and the commonly used GAI models in the Fashion design are illustrated in Section 5, highlighting the integration AI to enhance the Fashion design industry, and discussing promising approaches. Refer to figures 17-20 |
| **Q3.** | **How can the metaverse be used to improve the fashion industry?** | Section 6 presents an explanation of the metaverse and its evolution. Subsection 6.1 illustrates the use of the metaverse in the fashion industry. Section 6.2 demonstrates the purpose of AI in the metaverse for the fashion industry. Additionally, Subsection 6.3 presents caste studies of using the metaverse by fashion luxury brands. |
| Q4 | **What are the points of strength, weaknesses, opportunities, and threats of using the GAI in the Fashion industry?** | Subsection 7.1 presents a SWOT analysis for the use of GAI in the fashion industry. Through their points of strengths, GAI can help in inspiring fashion designers in addition to save time, effort and money while producing attractive designs. Weakness points have been explained so as to find a solution for them. |
| Q5 | **What are the points of strength, weaknesses, opportunities, and threats of using the metaverse in the Fashion industry?** | Subsection 7.2 presents a SWOT analysis for the use of the metaverse in the fashion industry. Metaverse allows for exposure and the ability to reach various users all over the world, show them different products through different events. Moreover, it is rich in data. These points of strength can benefit the fashion industry |
| Q6 | **How can GAI and the metaverse be integrated for fashion industry evolution?** | Subsection 7.3 proposes the idea of integration pointing out the benefits and the requirements that need to be considered when applying integration to achieve success. Use cases resulting from the integration are demonstrated. The idea of integration comes from the SWOT analyses of both GAI and the metaverse. The integration alleviates some of their weaknesses and allow more opportunities for the Fashion industry. Refer to Figure 21 |

Figure 19 depicts the percentages of the applications in which GAI have been used based on the covered references in this SLR. It shows that most of the application is garment generation followed by garment matching.

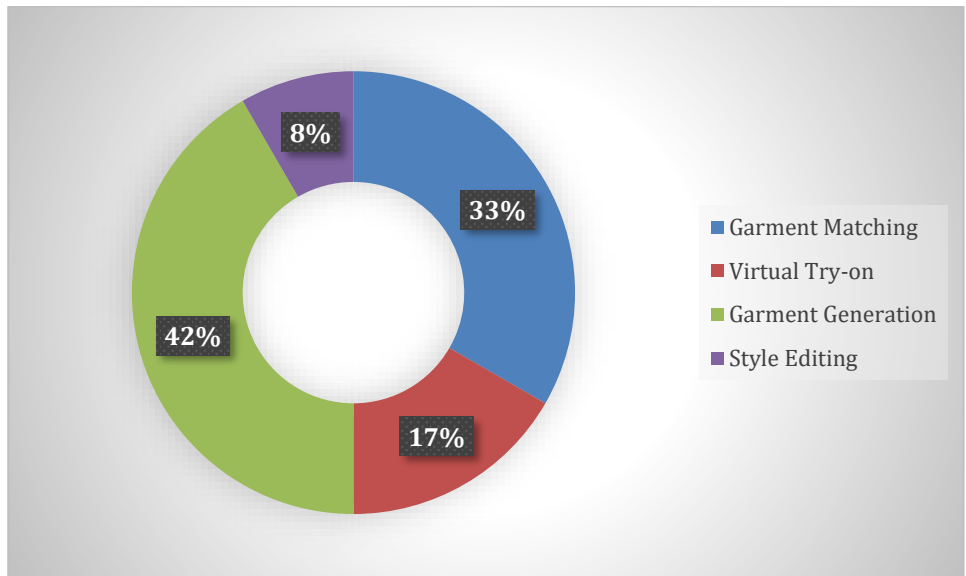

Fig. 19. The percentages of the applications in which GAI are used.

Figure 20 shows the number of times each dataset is used according to the references covered in this SLR. It is shown that most of the papers use collected data either from web scraping or collected from different datasets. Most of the datasets are collected from the Polyvore, which is a fashion social network allowing users to create fashion outfits from different items and DeepFashion. We included any collected dataset given a name in their references such as iFashion, MADAME and DiverseOut. It is noted that there is no one benchmark dataset that can be used for evaluation. Instead, each paper collects the suitable data for its purpose, which represents a challenge when comparing different techniques.

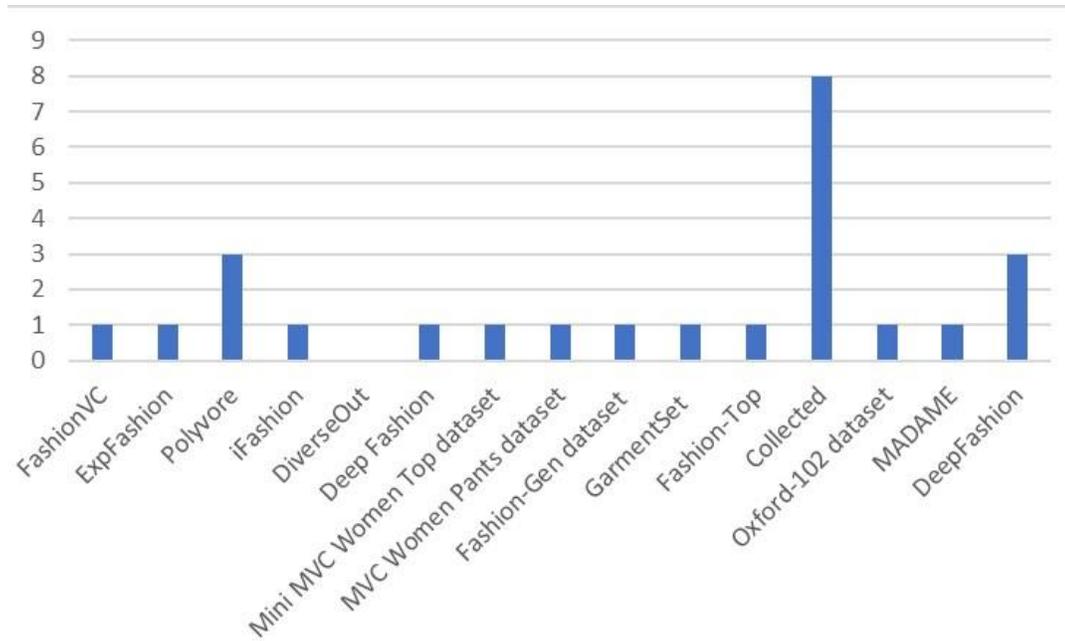

Fig. 20. The number of times each dataset is used.

Figure 21 demonstrates the types of GANs used in the references covered by this SLR and the number of times each type is used. It is noticed that most of the presented models are based on cGANs.

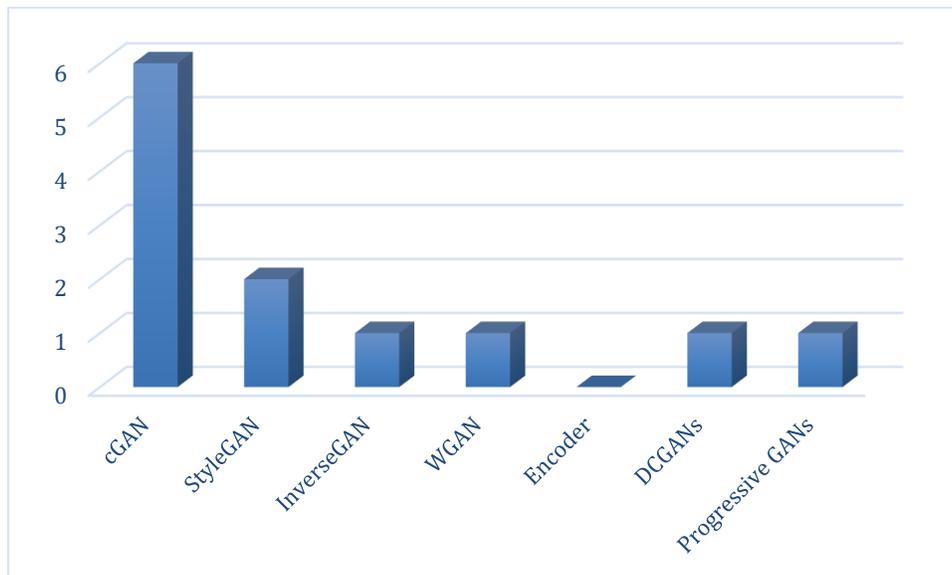

Fig. 21. The number of times each GAN type is used as a basic model in the references.

Figure 22 demonstrates the different performance measures used for evaluating the techniques of the papers covered in this SLR with the number of times each measure is used. It is shown that Human evaluation and FID are the most used measures. IS score is also frequently used.

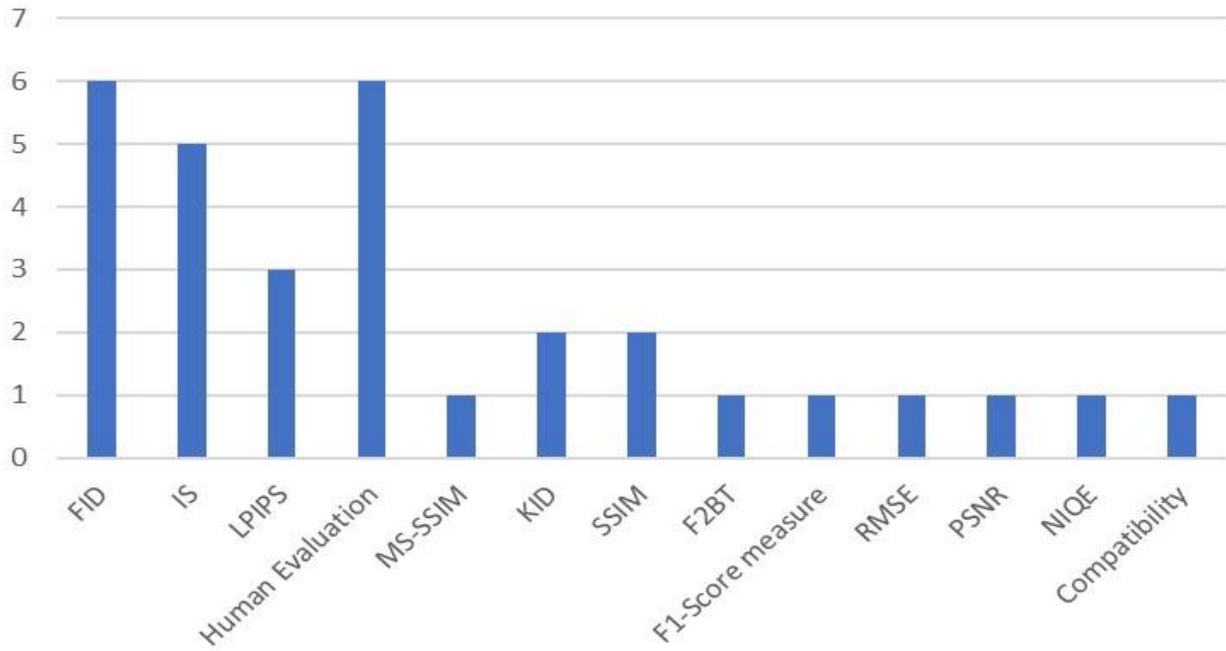

Fig. 22. The number of times each performance measure is used.

Figure 23 depicts the percentages of the main pillars used in answering the proposed questions. It is shown that discussing integration has the maximum percentage. This is because it is the main focus of this SLR.

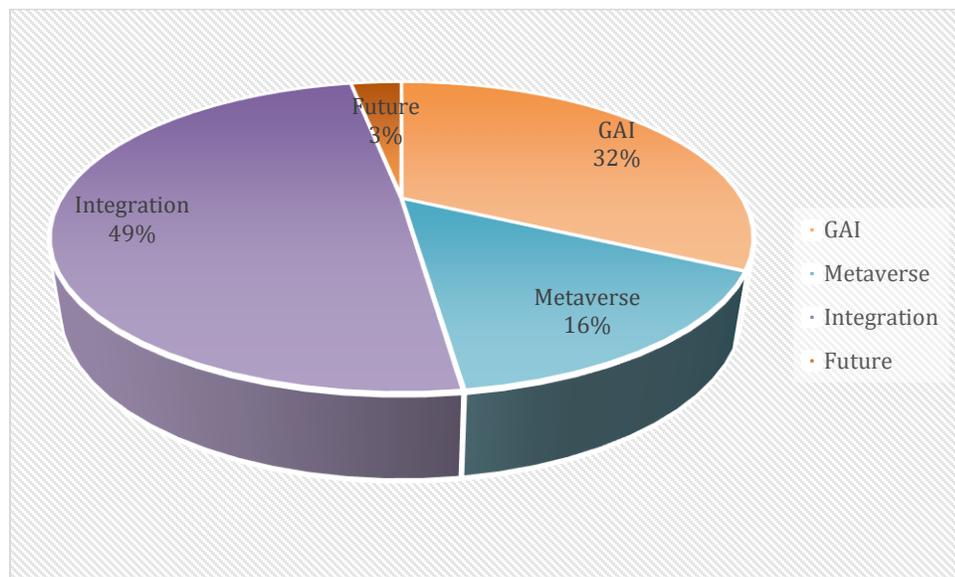

Fig. 23. Percentages of the main pillars of this SLR.

# 9 Conclusion and future directions

This SLR paper proposes the idea of integrating GAI and the metaverse to improve the Fashion industry. It focuses on how GAI can help fashion designers alleviate low-level and time-consuming tasks and create innovative designs. In addition, it explains how metaverse can be used to flourish marketing for different brands. It also provides a roadmap for researchers interested in using digital technologies of GAI and the metaverse. SWOT analyses have been performed showing the strengths and weaknesses of each technology in addition to the opportunities and threats of each. The benefits as well as the requirements of the integration have been explained thoroughly to provide a roadmap to achieve it. Moreover, future directions for researchers have been explained.

This SLR explores 188 papers including journal and conference articles by Web of Science, ACM Digital Library, Scopus, Science Direct, Taylor & Francis and JSTOR as the most promising scientific research database over the last ten years. Analytical statistics and discussion are presented.

**As for the future directions, may be the following points are changeable and are good direction as open research.**

1. For GAI: GAI can simplify the content creation process that can transform photographs into virtual reproductions. Various deep learning-based techniques can achieve high accuracy and real-time processing for displaying 3D objects. Researchers need to continue their working on developing new GAI techniques that support the fashion industry in terms of making photorealistic designs with an optimized performance. Designing performance measurement criteria is required to test the performance of different techniques.

2. For the metaverse: Researchers need to delve into the different methodologies to build immersive metaverse worlds using GAI. More specifically, creating avatars with different fashion designs, organizing fashion shows and creating whole innovative environments. Several hardware devices can be used to engage the user with the immersive experience such as VR devices.

3. Open research areas: Creating personalized designs is an open research area that involves monitoring user tastes and using GAI to generate similar designs. Disentanglement is still not widely used in the Fashion industry, yet, when it is used it boosts the performance. Also, InverseGANs are still not used much although using them may alleviate the need to use paired data. Different methods should be investigated to reduce the need for paired or annotated data. Benchmark datasets with varied features should be created to simplify the comparison between different techniques. Security and data privacy techniques need to be developed for both the metaverse and the use of GAI.

**Declarations**

**Conflict of interest:** There is no conflict of Interest.

**Ethical considerations:** There are no animal or human subjects in this paper.

**Funding:** No funding received.

**Consent for publication:** Agree to publish.

**Availability of data and materials:** There is no data sets used in this paper.